\documentclass{aa}  
\usepackage[varg]{txfonts}
\usepackage{rotating}
\usepackage{ulem}
\usepackage{xcolor}
\usepackage[breaklinks, colorlinks, citecolor=blue, urlcolor = blue]{hyperref}      
\usepackage{csquotes}
\usepackage{units}
\usepackage{mathtools}
\usepackage{float}
\usepackage[caption = false]{subfig}
\usepackage{graphicx}
\usepackage{amsmath}
\usepackage{footmisc}
\usepackage{multirow}

\bibpunct{(}{)}{;}{a}{}{,} 

\newcommand \sw{{\it Swift}}
\newcommand \fe{{\it Fermi}}
\newcommand \eb{{$E_{\rm break}$}}
\newcommand \ep{{$E_{\rm peak}$}}
\newcommand{\order}[1]{} %to impose the correct order in reference with same first author and same year
\begin{document} 

\title{Evidence of two spectral breaks in the prompt emission of gamma-ray bursts}
\titlerunning{spectral breaks in GRB prompt emission}
\author{M.~E. Ravasio\inst{\ref{inst1},\ref{inst2}}
\and G. Ghirlanda\inst{\ref{inst2},\ref{inst3},\ref{inst1}}
\and L. Nava\inst{\ref{inst2},\ref{inst4},\ref{inst5}}
\and G. Ghisellini\inst{\ref{inst2}}
}

\institute{ Universit\`a degli Studi di Milano-Bicocca, Dipartimento di Fisica U2, Piazza della Scienza, 3, I--20126, Milano, Italy \\
\email{\href{m.ravasio5@campus.unimib.it}{m.ravasio5@campus.unimib.it}}\label{inst1}
\and
INAF -- Osservatorio Astronomico di Brera, via Bianchi 46, I--23807 Merate (LC), Italy \label{inst2}
\and
INFN -- Milano Bicocca, Piazza della Scienza 3, I--20123, Milano, Italy \label{inst3}
\and
INAF -- Osservatorio Astronomico di Trieste, via G.B. Tiepolo 11, I--34143 Trieste, Italy \label{inst4}
\and
INFN -- via Valerio 2, I-34127 Trieste, Italy \label{inst5}
}  
\date{Received ; accepted}

\abstract
{The long-lasting tension between the observed spectra of gamma-ray bursts (GRBs) and the predicted synchrotron emission spectrum might be solved if electrons do not completely cool. Evidence of incomplete cooling was recently found in \sw\ GRBs with prompt observations down to 0.1\,keV, and in one bright \fe\ burst, GRB\,160625B. 
Here we systematically search for evidence of incomplete cooling in the spectra of the ten brightest short and long GRBs observed by \fe. We find that in eight out of ten long GRBs there is compelling evidence of a low-energy break (below the peak energy) and good agreement with the photon indices of the synchrotron spectrum (respectively -2/3 and -3/2  below the break and between the break and the peak energy). Interestingly, none of the ten short GRBs analysed shows a break, but the low-energy spectral slope is consistent with -2/3. In a standard scenario, these results imply a very low magnetic field in the emission region ($B^\prime \sim 10$\,G in the comoving frame), at odd with expectations. 
}


\maketitle

%============================ INTRODUCTION ============================
\section{Introduction}

The nature of the mechanism responsible for the hard X-ray and $\gamma$-ray prompt emission in gamma-ray bursts (GRBs) has been for years the subject of an intense debate and is still uncertain. 
Synchrotron emission has been proposed as the most natural radiative process, due to the non-thermal appearance of the observed spectra and to the likely presence of accelerated electrons and intense magnetic fields \citep{Rees94,Katz94,Tavani96,Sari96,Sari98}.
The debate is based on the inconsistency between the thousands of GRB spectra detected by different instruments and the spectral shape expected for synchrotron emission. 

The observed GRB prompt spectrum is indeed typically satisfactorily fitted by a smoothly broken power-law function, with photon flux described by $N(E) \propto E^{\alpha}$ at low energies and by $N(E) \propto E^{\beta}$ at high energies. The transition is smooth and identifies a typical break energy, which is the peak energy, \ep, in the $\nu F_{\nu}$ spectral representation. Typical values derived for long GRBs are $\alpha$\,$\sim$\,--1, $\beta$\,$\sim$\,--2.5, and \ep\,$\sim$\,200\,keV. This has been confirmed by the analysis of large samples of GRBs detected by the Burst And Transient Source Experiment (BATSE,  $\sim$\,20\,keV -- 2\,MeV; \citealt{Preece2000}, \citealt{Kaneko2006}) and the Gamma Burst Monitor (GBM, $\sim$\,8\,keV -- 40\,MeV, \citealt{Nava2011,Goldstein12,Gruber2014}).
Prompt emission spectra of short GRBs appear in general harder, not only in terms of peak energy ($E_{\rm peak}^{\rm short}$\,$\sim$\,0.5--1\,MeV), but also in terms of photon index $\alpha$. 
\cite{Ghirlanda09}
found that the low-energy spectral index of short bursts detected by BATSE has an average value $\alpha_{\rm short} = -0.4 \pm 0.5$. These results were later confirmed, also by GBM data \citep{Nava2011}.

The values of the low-energy photon index inferred from the observed spectra are in contrast with the predictions from the synchrotron theory. In the case of efficient cooling of the non-thermal population of electrons \citep{Sari98, Ghisellini2000} the predicted photon index is $\alpha_2^{\rm syn}$ = --3/2, significantly softer than the observed value.
A harder photon index ($\alpha^{\rm syn}_1$ = --2/3) is expected to describe the spectrum only at very low frequencies, below the cooling frequency.  
However, a small but sizable fraction of GRBs has been found to violate this 
limit \citep{Preece1998}, having photon index $\alpha > -2/3$.
These inconsistencies have been the major arguments against the synchrotron process for many years.

A few theoretical models have been proposed to reconcile the observed GRB prompt spectra with the synchrotron process. Some of them invoke effects that produce a hardening of the low-energy spectral index, such as a decaying magnetic field \citep{Peer2006,Uhm2014}, inverse Compton scattering in the Klein--Nishina regime, or a marginally fast cooling regime \citep{Derishev01,Nakar09,Daigne11}.

The advantages and difficulties of these and other models have been recently reviewed by \cite{Kumar2015}. These theoretical efforts have tried to modify the models in order to reproduce a typical photon index of $\alpha=-1$. Only recently has the problem  been tackled from the opposite side, through a revision of the way spectra can be modelled.

\citealt{Zheng2012}, analysing the X-ray and  $\gamma$-ray emission of GRB 110205A, as detected by \textit{Swift} and \textit{Suzaku}, identified a low-energy break in the prompt spectrum whose shape agrees with the synchrotron model.
A major advancement in the systematic characterisation of the low-energy part of prompt spectra has been made in two recent studies by \cite{gor2017a, gor2018}. 
They considered a sample of 34 long GRBs with prompt emission detected simultaneously by the Burst and Alert Telescope (BAT; 15--150\,keV) and by the X-Ray Telescope (XRT; 0.3--10\,keV) on board the {\it Swift} satellite. The joint spectral analysis revealed the presence (in most of the spectra) of a spectral break at low energies, around $\sim 2-30$ keV, in addition to the typical break corresponding to the peak energy. 
Remarkably, the two power-law photon indices $\alpha_1$ and $\alpha_2$, describing the spectrum below and above the newly found break energy, have distributions  centred around $-2/3$ and $-3/2$, respectively, consistent with the expectations of synchrotron theory.
The same spectral shape was found in GRB\,160625B \citep{Ravasio2018}, one of the brightest bursts detected by the \fe/GBM. Both the time-integrated and time-resolved spectra of this burst are characterised by a low-energy power-law photon index consistent with $\alpha_1^{\rm syn}=-2/3$, a spectral break at $\sim$\,50--100\,keV, a second power-law  photon index consistent with $\alpha_2^{\rm syn}=-3/2$ at intermediate energies, a second spectral break (representing the peak in $\nu F_\nu$) varying with time in the range \ep\,$\sim$\,300\,keV--6\,MeV, and a third power-law segment $\beta$\,$\sim$\,--2.6 describing the spectrum above \ep. 

These results triggered deeper investigations on the consistency of the spectra with synchrotron emission in a marginally fast cooling regime (i.e. with cooling frequency $\nu_{\rm c}$ smaller but comparable with the characteristic frequency $\nu_{\rm m}$: $\nu_{\rm c}\lesssim\nu_{\rm m}$). 
Oganesyan et al. (2019, in preparation) performed spectral fitting using a synchrotron model instead of empirical functions and testing the low-energy spectral shape thanks to the inclusion of simultaneous optical detections, concluding that the synchrotron spectral shape is a good fit to the data and that the optical flux lies on the extrapolation of the synchrotron spectrum.
A synchrotron model with $\nu_{\rm c}\sim\nu_{\rm m}$ was also found to be a good description of the data  in 19 single-pulse GBM bursts \citep{Burgess2018}.

In this paper, motivated by the identification of two spectral breaks in the \fe\ burst GRB\,160625B \citep{Ravasio2018}, we report on the systematic search for this feature in the brightest \fe/GBM bursts. 
To date, the presence of a low-energy break has been reported only in long GRBs \citep{gor2017a,gor2018,Ravasio2018}. 
We extended, for the first time, the search for this feature to  short GRBs as well.
We selected the ten brightest long GRBs and the ten brightest short GRBs detected 
by the GBM (\S\ref{sec:sample}). 
We performed a spectral analysis to identify the possible presence of a low-energy spectral break, following the method described in \S\ref{sect:data_analysis}. 
The results are presented in \S\ref{sect:results}, and a discussion of their physical implications in the context of the GRB standard model is proposed in \S\ref{sect:discussion}. 
In \S\ref{sect:conclusions} we summarise the main results of this work.

%============================ SAMPLE SELECTION and properties ============================
\section{The sample}\label{sec:sample}

We sorted the GRBs included in the online GBM Catalogue\footnote{https://heasarc.gsfc.nasa.gov/W3Browse/fermi/fermigbrst.html \label{fn:GBMcatalog}} according to their 10--1000\,keV fluence of the best fitting model and selected the brightest  ten from   the long class and ten from the short class. 
This selection corresponds to fluence cuts $F > 1.79 \times 10^{-4}\,\rm erg\,cm^{-2}$ and  $F > 5.72\times10^{-6}\,\rm erg\,cm^{-2}$ for long and short GRBs, respectively. The list of selected events is reported in Table~\ref{tab:longGRB} (long GRBs) and in Table~\ref{tab:shortGRB} (short GRBs). 

A selection based on the fluence ensures a good photon statistics (required to identify, with a certain degree of confidence, a possible low-energy break) and the possibility of performing time-resolved analysis.
This is crucial to study if and how this spectral feature evolves in time, and whether its evolution is related to other evolving quantities 
such as the peak energy.
From our selection we excluded GRB\,090902B and GRB\,130427A 
for the following reasons.
GRB\,090902B, which would satisfy our selection, has a prominent high-energy emission detected by the LAT during the prompt phase, which extends low energies and dominates the emission below $\sim$\,30\,keV \citep{090902B}. Moreover, as shown in \cite{Ryde2010} and \cite{Peer2012}, its spectrum seems to be dominated by a thermal photospheric emission component. These reasons both prevent the identification of a possible low-energy break in the main spectral component, which is the feature we want to investigate in this work.
GRB\,130427A, due to its large fluence, suffered from pile-up effects \citep{Preece2014} and a standard analysis can be performed only on its precursor, which does not satisfy our selection criterion. We thus excluded this GRB from our sample. We note, however, that a spectral analysis of the precursor is reported by \cite{Preece2014}, who find consistency with synchrotron emission. 
One of the GRBs included in our sample, GRB 160625B, was already analysed in \citet{Ravasio2018}, who identified a clear spectral break and a good consistency of the overall spectrum with synchrotron radiation in a marginally fast cooling regime (see also \citealt{Zhang2018, Wang2017, Lu2017}).
For homogeneity, here we reanalyse its spectra with the same procedure adopted in this work for the other bursts.

%----------------------- table long -------------
\begin{table}
\centering 
\caption{Ten long GRBs with the largest fluence (10--1000\,keV) in the GBM catalogue. 
The last three digits in the name (in square brackets) refer to the naming convention of GBM triggers.
The prompt duration and the 10--1000\,keV fluence of the time-integrated spectra are listed in Cols. 2 and 3, and refer to information reported in the online GBM catalogue. The last column gives the redshift, if available.}
\label{tab:longGRB}
\tiny
\begin{tabular}{cccc}
\hline\hline
 \multicolumn{1}{c}{GRB Name} &
 \multicolumn{1}{c}{$T_{90}$} &
 \multicolumn{1}{c}{Fluence} &
 \multicolumn{1}{c}{Redshift}\\
  & [s] & [$10^{-4}$\,erg\,cm$^{-2}$] &\\
\hline
171010[792] & $ 107.27 \pm 0.81 $ & $ 6.72 \pm 0.02 $  & 0.3285 \\
160625[945] & $ 453.38 \pm 0.57 $ & $ 6.68 \pm 0.02 $  & 1.406 \\
160821[857] & $ 43.01 \pm 0.72 $ & $ 5.48 \pm 0.02 $  & -- \\
170409[112] & $ 64.0 \pm 0.72 $ & $ 3.19 \pm 0.01 $  & -- \\
180720[598] & $ 48.9 \pm 0.36 $ & $ 3.18 \pm 0.01 $  & 0.654 \\
171227[000] & $ 37.63 \pm 0.57 $ & $ 3.05 \pm 0.01 $  & -- \\
090618[353] & $ 112.39 \pm 1.09 $ & $ 2.74 \pm 0.02 $  & 0.54 \\
100724[029] & $ 114.69 \pm 3.24 $ & $ 2.43 \pm 0.01 $  & -- \\
130606[497] & $ 52.22 \pm 0.72 $ & $ 2.15 \pm 0.01 $  & -- \\
101014[175] & $ 449.42 \pm 1.41 $ & $ 1.79 \pm 0.01 $  & -- \\
\hline
\end{tabular}
\end{table}
%----------------------- end table long ------------------------------
%-------------------------- table short -----------------------------
\begin{table}
\centering 
\caption{Ten short GRBs with the largest fluence (10--1000\,keV) in the GBM catalogue. 
The last three digits in the name (in square brackets) refer to the naming convention of GBM triggers.
The prompt duration and the fluence of the time-integrated spectra are listed in Cols. 2 and 3, and refer to information reported in the online GBM catalogue.}
\label{tab:shortGRB}
\small
\begin{tabular}{ccccc}
\hline\hline
 \multicolumn{1}{c}{GRB Name} &
 \multicolumn{1}{c}{$T_{90}$} &
 \multicolumn{1}{c}{Fluence} \\
 & [s] & [$10^{-6}$\,erg\,cm$^{-2}$]\\
\hline
170206[453] & $ 1.17 \pm 0.10 $ & $ 10.80 \pm 0.16 $ \\
120323[507] & $ 0.38 \pm 0.04 $ & $ 10.66 \pm 0.13 $ \\
140209[313] & $ 1.41 \pm 0.26 $ & $ 9.52 \pm 0.18 $ \\
090227[772] & $ 0.30 \pm 0.02 $ & $ 8.93 \pm 0.17 $ \\
150819[440] & $ 0.96 \pm 0.09 $ & $ 7.75 \pm 0.15 $ \\
170127[067] & $ 0.13 \pm 0.04 $ & $ 7.41 \pm 0.21 $ \\
120624[309] & $ 0.64 \pm 0.16 $ & $ 7.14 \pm 0.16 $ \\
130701[761] & $ 1.60 \pm 0.14 $ & $ 6.30 \pm 0.14 $ \\
130504[314] & $ 0.38 \pm 0.18 $ & $ 6.01 \pm 0.14 $ \\
090228[204] & $ 0.45 \pm 0.14 $ & $ 5.72 \pm 0.11 $ \\
\hline
\end{tabular}
\end{table}
%---------------------------- end table short ------------------------

%======================================= DATA ANALYSIS ====================================
\section{Spectral analysis}\label{sect:data_analysis}

The GBM is composed of 12 sodium iodide (NaI, 8\,keV--1\,MeV) and 2 bismuth germanate (BGO, 200\,keV to 40\,MeV) scintillation detectors \citep{Meegan2009}. We analysed the data from the two NaI and one BGO with the highest count rate.
For long GRBs we used CSPEC data, which have 1024\,ms time resolution, while for short GRBs we selected Time Tagged Event (TTE) data, with shorter time binning (64\,ms). Spectral data files and the corresponding response files were obtained from the online archive\footref{fn:GBMcatalog}. 
Spectral analysis was performed with the public software {\sc rmfit}~(v.~4.3.2).
We followed the procedure explained in the Data Analysis Threads and Caveats\footnote{https://fermi.gsfc.nasa.gov/ssc/data/}.
In particular, we selected the energy channels in the range 8--900\,keV for NaI detectors, and 0.3--40\,MeV for BGO detectors, and excluded the channels in the range 30--40\,keV due to the presence of the iodine K-edge at 33.17\,keV\footnote{https://fermi.gsfc.nasa.gov/ssc/data/analysis/GBM\_caveats.html}.
To model the background, we selected background spectra in time intervals before and after the burst
and modelled them with a polynomial function up to the fourth order.
For the time-resolved analysis, the light curve is rebinned imposing a signal-to-noise ratio $S/N>60$. 
This selection is performed on the most illuminated NaI detector. 
The choice of optimising the $S/N$ of the NaI, regardless of the $S/N$ in the BGO, is motivated by our interest in the low-energy break, that (if present) lies below 100\,keV, i.e. within the energy range of the NaI detectors 
(8 -- 900\,keV). 
Given the relatively large value of the $S/N$, the $\chi^2$ statistic is used in the fitting procedure.

We analysed both time-integrated and time-resolved spectra with two different empirical functions: a smoothly broken power law (SBPL) and a double smoothly broken power law (2SBPL; see \citealt{Ravasio2018} for the description of their functional form). 
The SBPL is made of two power laws, with spectral indices $\alpha$ and $\beta$, smoothly connected at some break energy (usually corresponding to the $\nu F_\nu$ peak of the spectrum, \ep). 
The 2SBPL is a single continuous function that allows  the spectra to be fit with three power laws (with photon indices named $\alpha_1$, $\alpha_2$, and $\beta$) smoothly connected at two breaks (hereafter $E_{\rm break}$ and $E_{\rm peak}$). The 2SBPL function was found to fit the spectrum of GRB 160625B significantly better than the SBPL, revealing the presence of a break at low energies in addition to the usual peak of the $\nu F_{\nu}$ spectrum. 
An example of a spectrum fitted with the 2SBPL function is shown in Fig.~\ref{fig:spettro180720}.
In particular it refers to the time bin 7.17 -- 8.19\,s of GRB\,180720. 
The best value for the low-energy break is $E_{\rm break} = 93.62_{-64.1}^{+91.6}$\,keV and the photon indices of the power law below and above it are 
$\alpha_1=-0.71_{-0.46}^{+0.13}$ and $\alpha_2 = -1.47_{-0.26}^{+0.20}$, 
while the peak energy is $E_{\rm peak} = 2.42_{-0.64}^{+1.02}$\,MeV and the high-energy photon index $\beta = -2.38_{-0.30}^{+0.23}$.
The plot also shows  for comparison the 
power laws expected from synchrotron emission (dashed lines).

The SBPL is one of the empirical functions generally used to model GRB spectra \citep{Kaneko2006,Gruber2014}. 
It gives more flexibility than the Band function to properly model the curvature around \ep, at the expense of having one additional free parameter. However, introducing a fifth free parameter usually results in ill-determined unconstrained parameters and degeneracy or correlations among them \citep{Kaneko2006}. For this reason, the value describing the curvature is usually kept fixed to a value that has been found to satisfactorily describe most of the spectra \citep{Goldstein12,Gruber2014}.
The problem is even more severe when fitting a 2SBPL, which has eight free parameters, two of them describing the curvature around \eb\ and around \ep. 
We decided to fix the values of the parameters describing the curvatures, both for the SBPL and for the 2SBPL.
Since we wanted to test a synchrotron origin, we chose curvatures that reproduce the shape of synchrotron spectra. We built a synthetic synchrotron spectrum for a population of partially cooled electrons and compared it to the SBPL and 2SBPL to find for which values of the curvatures these empirical functions mimic the shape of the synchrotron spectrum. We repeated the test for different cooling efficiencies (i.e. for different values of the ratio \ep/\eb) and derived that the most suitable parameters for the curvature are $n=2$ (see \citealt{Ravasio2018} for their definition).
These values correspond to very smooth curvatures.

For the joint analysis of the two NaI and the BGO data we used an intercalibration constant factor between the brightest NaI and the other NaI and BGO detectors.
Since we were comparing two nested models, the best fit model for each analysed spectrum was chosen by applying a $\chi^{2}$-based $F$-test: we select the more complex model (2SBPL) only if it corresponds to an improvement with a significance larger than $3\sigma$.

%======================================= RESULTS ======================================
\section{Results}\label{sect:results}

In this section, we present separately the results of the spectral analysis for the samples of long and short GRBs listed in Table~\ref{tab:longGRB} and Table~\ref{tab:shortGRB}, respectively.

\subsection{Long GRBs}\label{subsec:long}
%-------------------------------- LONG -------------------------------
The results of the time-integrated analysis for the sample of long GRBs are given in Table~\ref{tab:params_timeint_long}. For each GRB, the table lists the GRB name  (in bold  if the best fit model is a 2SBPL, i.e. if a break is present), the time interval used for the time-integrated spectral analysis, the best fit parameters of the best fit model (either a SBPL or a 2SBPL) chosen according to the significance of the $F$-test (last column), and the total $\chi^2$/dof.

According to the $F$-test, in two long GRBs the improvement in the $\chi^2$ caused by the inclusion of a low-energy break in the fitting function has a significance corresponding to $1\sigma$ and $2\sigma$: in these two cases the best fit model is then a SBPL. 
On the contrary, in all   eight of the remaining long GRBs the 2SBPL function significantly 
improves the fit provided by the SBPL model (at more than 3$\sigma$). 
In particular, the improvement is significant at more than 8$\sigma$ in six cases and between 4$\sigma$ and 8$\sigma$ in two cases.
This means that in eight of the ten brightest long GRBs the time-integrated spectrum shows the presence of two characteristic energies: a low-energy spectral break \eb\ (typically between a few dozen and a few hundred  keV) and the usual peak of the $\nu F_\nu$ spectrum \ep\ (typically between a few hundred and a few thousand  keV). 
Table \ref{tab:params_timeint_long} lists the parameters of the best fit model for the time-integrated spectra of each long GRB.

In three cases where the presence of the low-energy spectral break is strongly supported by the significance of the $F$-test (namely, GRB\,171010, GRB\,090618, and GRB\,101014), the break energy is located at \eb\,$\sim$\,10\,keV, very close to the low-energy edge of the GBM ($\sim$\,8\,keV): very few data points are available below the break to properly constrain the value of photon index $\alpha_1$. 
In all of these cases, we find that the best fit value of $\alpha_1$ reaches very hard values, at odds with results derived when \eb\ is located at higher energies, far from the low-energy edge of the instrument. 
We discuss this issue in more detail in Appendix~\ref{appendix:Ebreaklow}. 
When calculating mean values we included only spectra with $E_{\rm break} > 20$\,keV. For the time-integrated analysis, the typical values of the parameters of the 2SBPL model are $\alpha_{1}=-0.76_{-0.03}^{+0.03}$, $\alpha_{2} = -1.65_{-0.04}^{+0.05}$, $\rm log(E_{\rm break})=2.11_{-1.08}^{+1.11}$, $\rm log(E_{\rm peak})=2.98_{-1.66}^{+1.72}$, and $\beta=-2.85_{-0.08}^{+0.09}$.

%--------------------------- LONG TIME-RESOLVED--------------------
For the eight long GRBs with a low-energy break, we also performed a time-resolved analysis to investigate the presence of the break on shorter timescales and its temporal evolution, also in comparison to the overall spectral evolution. 
The time-resolved spectral analysis is performed on temporal bins of 1.024\,s in width.  
We fit all the spectra with the SBPL and 2SBPL models. 
If the latter model results in a smaller $\chi^2$, we assess the significance of the improvement with the $F$-test. If there are two (or more) consecutive time bins where the 2SBPL does not produce a better fit  (i.e. with significance $<3\sigma$), we combine them  in order to acquire more statistics 
and further test the presence of a spectral break. 
This procedure shows that, in most cases, it is sufficient to combine two or three consecutive bins to constrain \eb.
This time-rebinning was applied to $\sim$\,28\% of the time-resolved spectra. 

All the results of the time-resolved analysis on the eight long GRBs that present a break energy \eb\ in the time-integrated spectra are summarised in Table~\ref{tab:params_timeres_long} and shown in Fig.~\ref{fig:timeevolutionpanels}.
For each GRB the upper panel shows the light curve, while the lower panels show the results of the spectral analysis: from top to bottom the spectral indices $\alpha_1$ and $\alpha_2$ (or $\alpha$ only, if the best fit model is a SBPL), the photon index $\beta$, the characteristic energies \eb\ and \ep\ (or \ep\ only, if the best fit model is a SBPL), and the ratio \ep/\eb. 
In most of the time-resolved spectra (139/199, i.e. $\sim$\,70\%) the best fit model is the 2SBPL function. 

%----------------------------- 
\begin{figure}[h]
\centering
\includegraphics[scale=0.16]{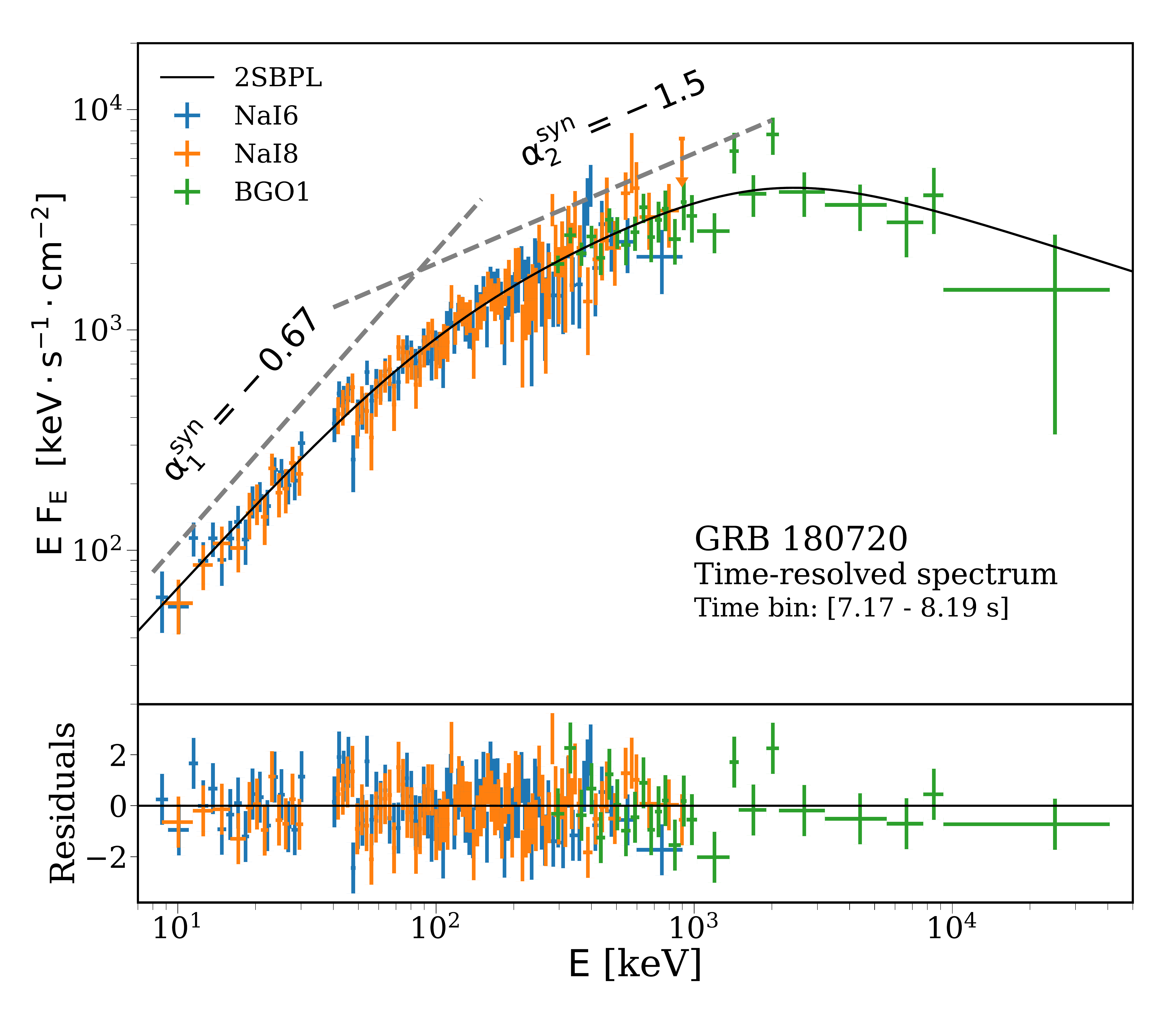} 
\caption{Example of a spectrum best fitted by a 2SBPL (i.e. three power laws smoothly connected at two breaks). The data correspond to the time interval 
7.17 -- 8.19\,s from the trigger of GRB\,180720. 
The best fit values of the 2SBPL model parameters are $\alpha_1=-0.71$, $E_{\rm break} = 93.62$\,keV, $\alpha_2 = -1.47$, $E_{\rm peak}= 2.42$\,MeV, and $\beta = -2.38$. The different instruments are colour-coded as shown in the legend. The two dashed lines show, for comparison, the power laws (with the photon indices) predicted by synchrotron emission. 
Data-to-model residuals are shown in the bottom panel.}
\label{fig:spettro180720}
\end{figure}
%----------------------------- 

%----------------------------- 
\begin{figure*}[!h]
\centering
\includegraphics[scale=0.5]{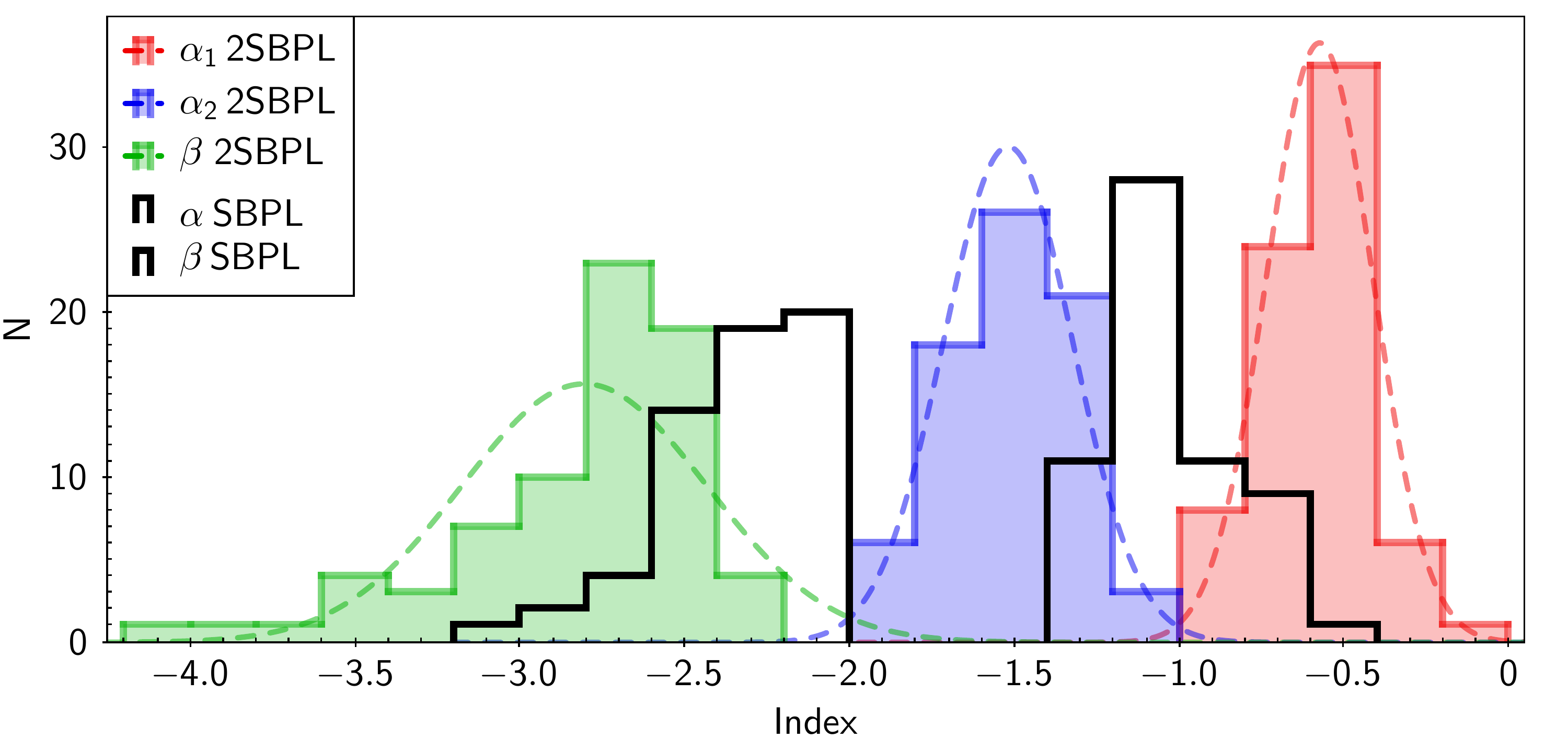} 
\caption{Long GRBs, time-resolved analysis: distribution of the spectral indices, according to the best fit model, for the time-resolved fits of the eight long GRBs showing a spectral break. 
The spectral indices $\alpha_1$, $\alpha_2$, and $\beta$ of the 2SBPL model are shown with red, blue, and green filled histograms, respectively. 
Gaussian functions showing the central value and standard deviation of the distributions are overlapped to the histograms (colour-coded dashed curves). The black empty histograms represent the distributions of the two photon indices $\alpha$ and $\beta$ of the SBPL model, for spectra where the SBPL is the best fit model.}
\label{fig:distrib_index_long_timeres}
\end{figure*}
%-----------------------------

%----------------------------- 
\begin{figure}[h]
\centering
\includegraphics[scale=0.47]{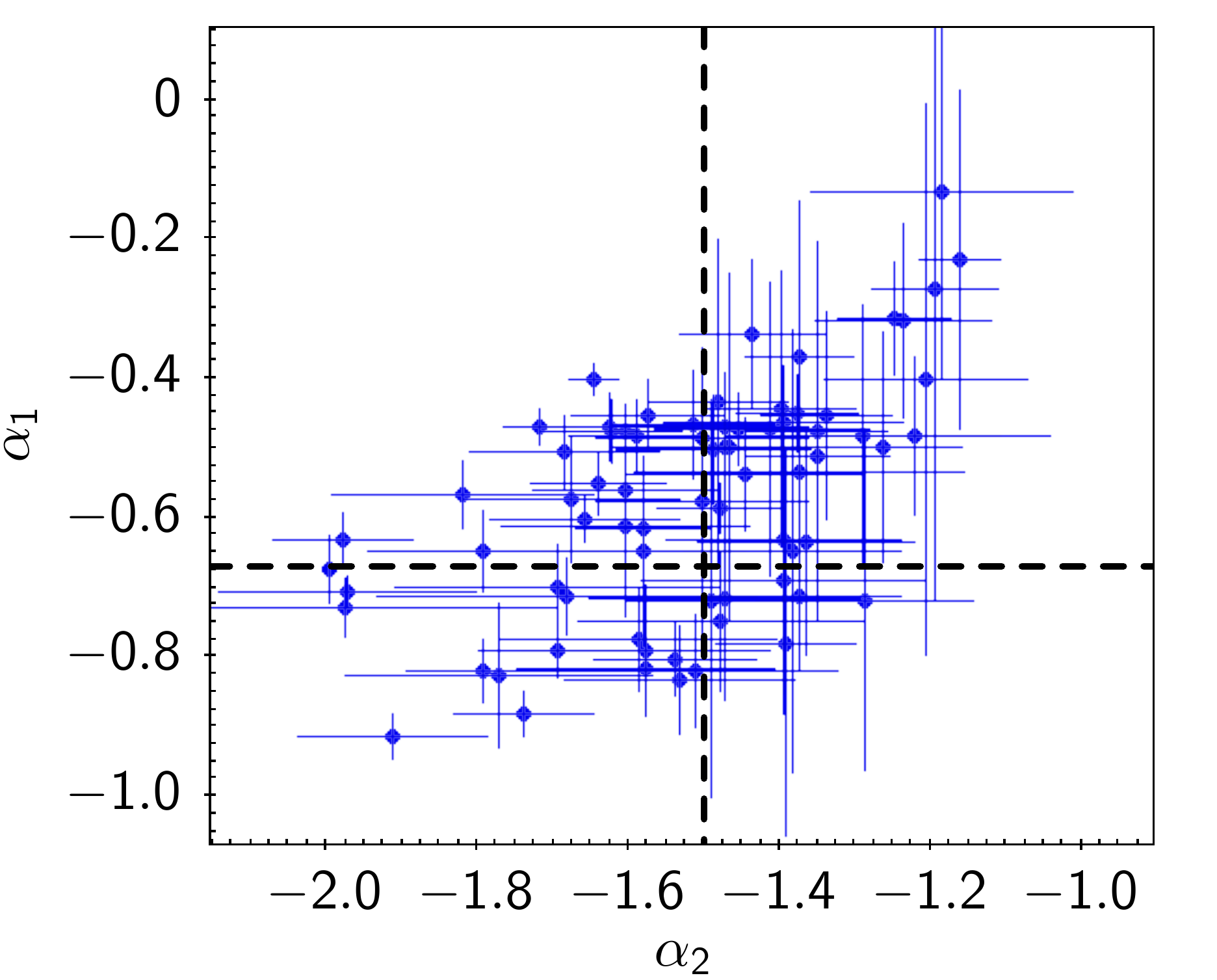} 
\caption{
Long GRBs, time-resolved analysis: power-law photon index $\alpha_1$ vs. the power-law photon index $\alpha_2$. 
The two black dashed lines mark the expected values for synchrotron emission.
}
\label{fig:index_long_timeres}
\end{figure}
%----------------------------- 

In the time-resolved spectra there are cases where \eb\ is close to the low-energy threshold of the \fe\ band (i.e. $\sim$\,10--20\,keV).
In particular, we find \eb$<20$\,keV in all time-resolved spectra of GRB 171010 and in 10 of the 35 time-resolved spectra of GRB 090618 and GRB 101014; it should be noted  that these three GRBs are the same that have \eb$<20$\,keV in the time-integrated spectrum).
As for the results of the time-integrated analysis, we consider in the following analysis the time-resolved spectra with $E_{\rm break}>20$\,keV
(see  Appendix~\ref{appendix:Ebreaklow} for a motivation of this choice). 

Figure~\ref{fig:distrib_index_long_timeres} shows the distribution of the spectral indices $\alpha_1$, $\alpha_2$, and $\beta$ of the 2SBPL model fits (filled histograms). 
These histograms are built considering the time-resolved spectra for which the 2SBPL is the best fit model. 
The inferred mean values are $\langle\alpha_1\rangle=-0.58$ (with standard deviation $\sigma_{\alpha_1}$ = 0.16) and $\langle\alpha_2\rangle=-1.52$ ($\sigma_{\alpha_2}$ = 0.20). 
These values are remarkably consistent with those predicted 
for a population of electrons emitting synchrotron radiation in the so-called fast cooling regime. 

For comparison, Fig.~\ref{fig:distrib_index_long_timeres} also shows the distributions (solid line, black histograms) of the spectral indices $\alpha$ and $\beta$ (i.e. below and above the peak energy \ep, respectively) for those spectra sufficiently well fitted by the SBPL (i.e. \eb\ is not required according to the $F$-test). 
The distribution of the spectral index $\alpha$ of the SBPL model is consistent with the value $\langle\alpha\rangle=-1.02$ ($\sigma_\alpha$ = 0.19) typically reported in the literature which is obtained employing single break fitting functions (e.g. SBPL or Band). It is interesting to note that this distribution is  placed almost in the middle of the two distributions of the spectral indices $\alpha_1$ and $\alpha_2$ of the 2SBPL model, namely of the two power laws below and above \eb.
As was done in \citealt{Ravasio2018}, we also performed the fit of the time resolved spectra by fixing the slope of the low-energy power-law index to the value $\alpha_1 =-2/3$ predicted by the synchrotron theory. We did this analysis for the spectra in which the low-energy power-law index is harder, at more than 1$\sigma$, than -2/3. These represent $47\%$ of the spectra. When we fix $\alpha_1$ =-2/3 in the 2SBPL model, most ($\sim 85\%$) of the time resolved spectra can still be adequately fitted (probability $> 10^{-2}$) and the other free parameters of the model assume values which are consistent, within their errors, with those obtained leaving $\alpha_1$ free.

The spectral index $\beta$, describing the high-energy part of the spectrum (i.e. above \ep), has a distribution centred around  $\langle\beta\rangle=-2.33$ ($\sigma_\beta = 0.24$) for the spectra fitted by the SBPL, while the 2SBPL fits provide a distribution centred at $\langle\beta\rangle=-2.81$ with ($\sigma_\beta=0.37$). 
Thus, when the spectrum requires the presence of two breaks (i.e. three power laws) the high-energy power law is steeper than the cases when only one break is present.

The scatter plot of $\alpha_1$ versus $\alpha_2$ is shown in Fig.~\ref{fig:index_long_timeres}. 
The reference synchrotron values are shown with dashed lines.  
Despite the large scatter of the data points, a correlation analysis suggests that a statistically significant correlation (with correlation coefficient $\rho=0.35$ and chance probability $P=0.002$) is present. 
Also, within individual GRBs (shown in Appendix~\ref{appendix:timeevolutionpanels}) the two indices seem to track each other.

\begin{figure}[t]
\centering
\includegraphics[scale=0.45]{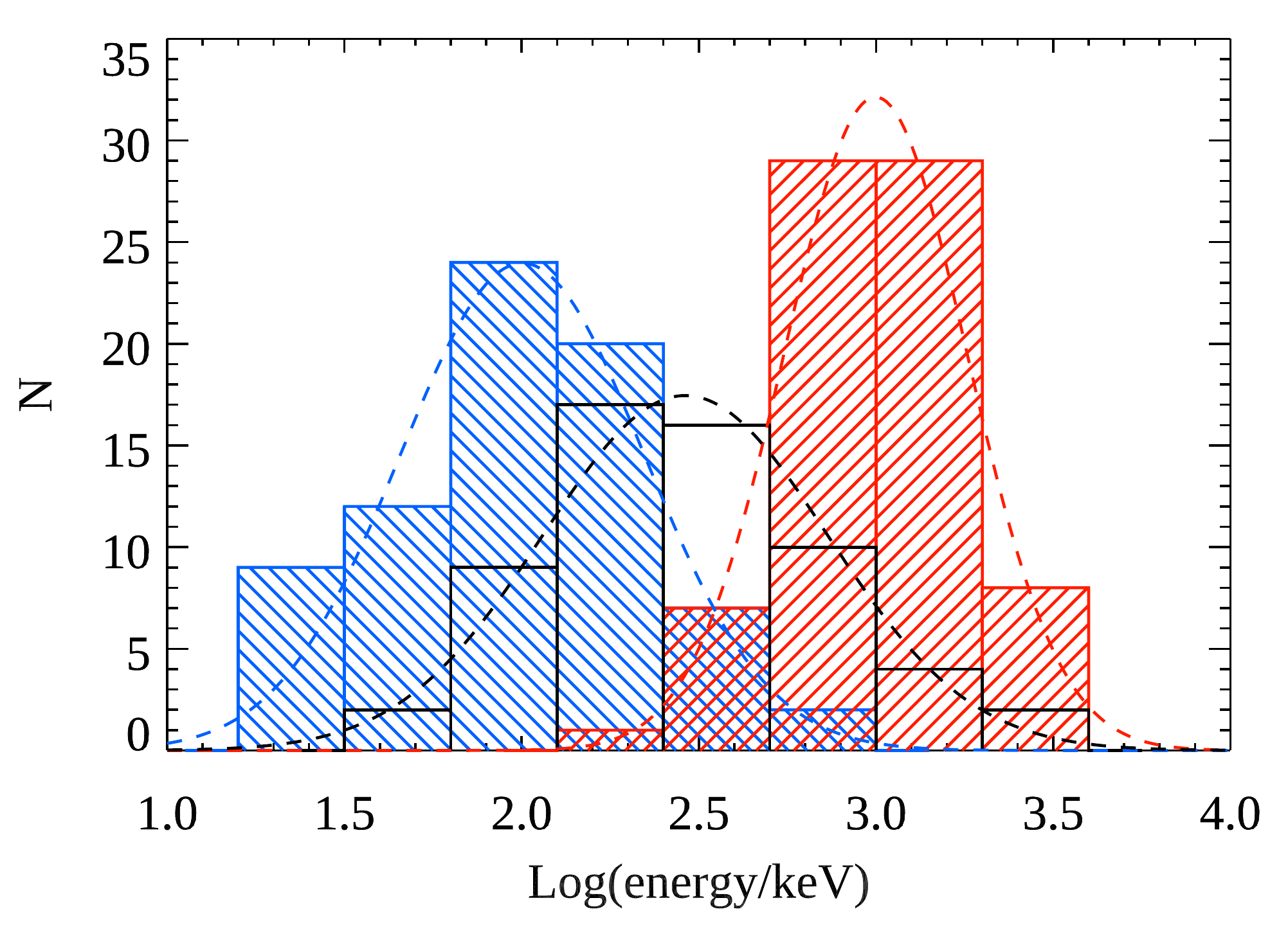}
\includegraphics[scale=0.42]{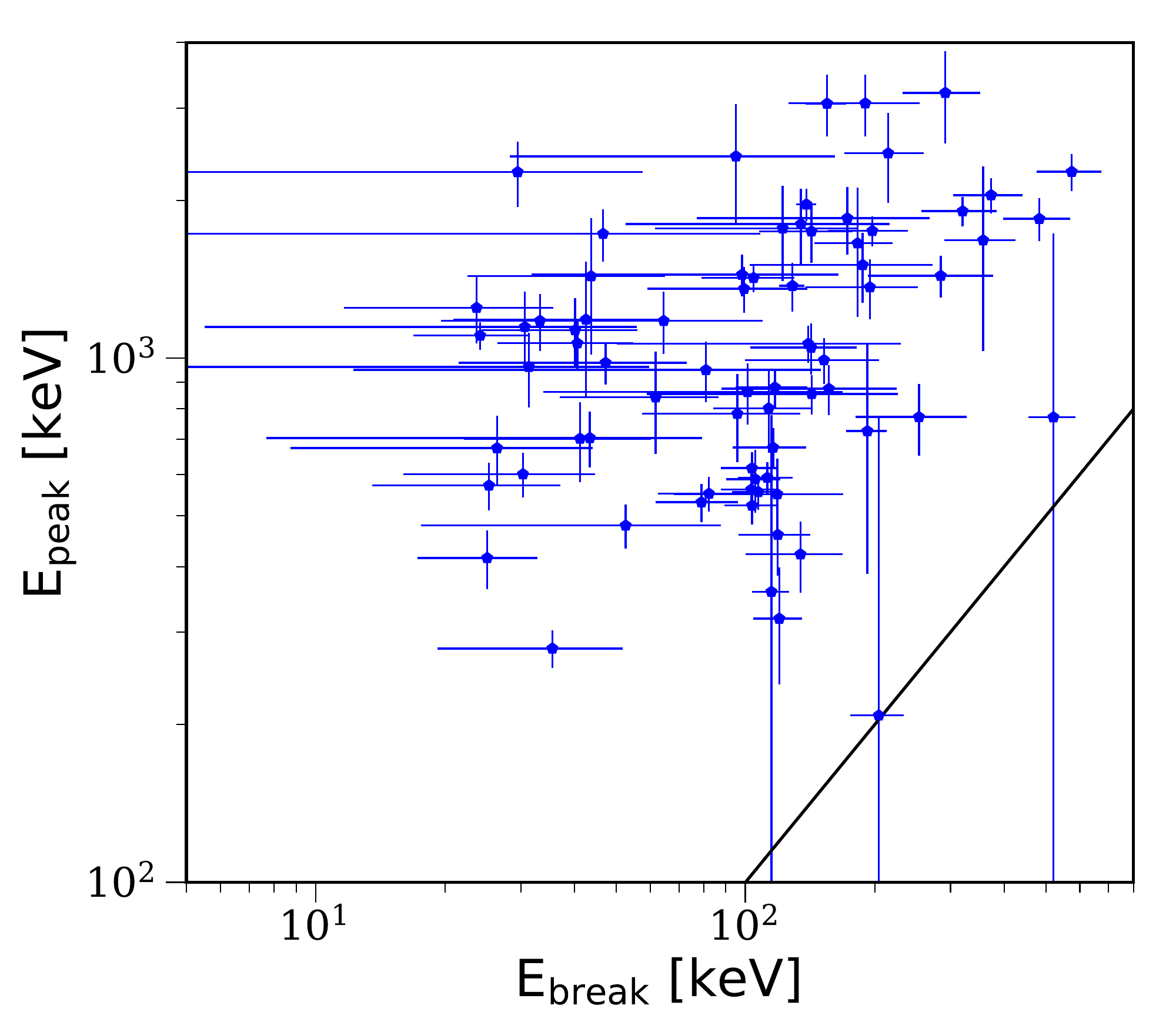} 
\caption{Long GRBs, time-resolved analysis. 
{\it Top panel:} Distributions of the characteristic energies ($E_{\rm beak}$ and $E_{\rm peak}$ for the 2SBPL (blue and red hatched histogram, respectively) and $E_{\rm peak}$ for the SBPL model (black empty histogram). Gaussian functions showing the central value and standard deviation for each distribution are overplotted to the histograms (with the same colour-coding). 
{\it Bottom panel:} $E_{\rm peak}$ versus $E_{\rm break}$  scatter plot (2SBPL model). The equality line is shown with a solid line.}
\label{fig:energies_long_timeres}
\end{figure}

The top panel of Figure~\ref{fig:energies_long_timeres} shows the distributions of the two characteristic energies of the 2SBPL fits (blue histograms for \eb\ and red histograms for \ep), and \ep\ of the SBPL fit (black empty histogram). $E_{\rm peak}^{\rm SBPL}$ has a log-normal distribution centred at $\langle \log(E_{\rm peak}^{\rm SBPL}\rm{/keV}) \rangle=2.46$ ($\sigma_{E_{peak}}$ = 0.40). Instead, when a second break in the fitting function is introduced and its presence in the spectrum is statistically significant, the distributions of $E_{\rm break}$ and $E_{\rm peak}$ are centred at the mean values $\langle \log(E_{\rm break}{\rm /keV})\rangle=2.00$ ($\sigma_{E_{break}}$ = 0.34) and $\langle \log(E_{\rm peak}^{\rm 2SBPL}{\rm /keV}) \rangle=3.00$ ($\sigma_{E_{peak}}$ = 0.26). 
The bottom panel shows the scatter plot of  $E_{\rm peak}$ and $E_{\rm break}$ obtained from the 2SBPL fits.

From the comparison of the best fit values obtained when the best fit model is a SBPL and when is a 2SBPL (Fig.~\ref{fig:distrib_index_long_timeres} and Fig. \ref{fig:energies_long_timeres}) we notice the following: i) the distribution of $\alpha^{\rm SBPL}$ lies  between the distributions of $\alpha_1$ and $\alpha_2$ (see also \citealt{gor2018}); ii) on average $E_{\rm peak}^{\rm SBPL}<E_{\rm peak}^{\rm 2SBPL}$; and iii) $\beta^{\rm SBPL}>\beta^{\rm 2SBPL}$.

%----------------------------- 
%----------------------------- 
%-------------------------------- SHORT -------------------------------
\subsection{Short GRBs}\label{subsec:short}
For the short GRBs in our sample we analyse only the time-integrated spectra because we find that there is not enough signal to separate them in several bins as we did for long GRBs.
The results for each GRB are shown in Table~\ref{tab:params_timeint_short}.

\begin{figure}[t]
\centering
\includegraphics[scale=0.45]{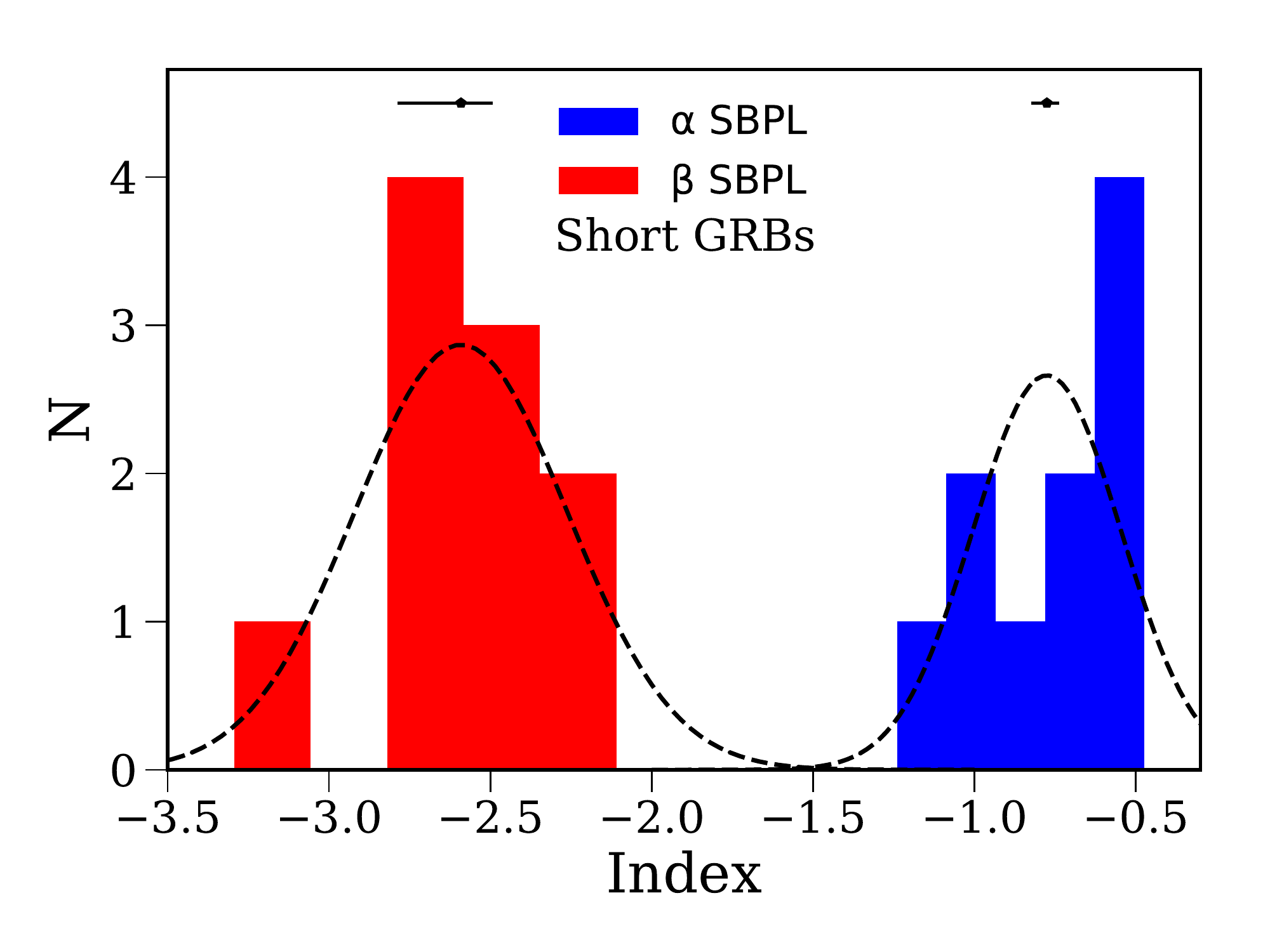}
\caption{Short GRBs, time-integrated spectra: distributions of the spectral indices $\alpha$ and $\beta$ for the best fit model (which is always a SBPL) for all ten short GRBs  in Table~\ref{tab:params_timeint_short}. The mean values and typical errors are shown at the top of the corresponding distributions (black symbols). Gaussian functions showing the mean value and standard deviation are overplotted on the histograms.
}\label{fig:index_short_timeint}
\end{figure}

Contrary to what was found in long GRBs, none of the ten short GRB time-integrated spectra shows evidence for a low-energy spectral break.
They are all well fitted by the SBPL function, thus by two power laws smoothly connected at the $\nu F_\nu$ peak.
This peak energy has a typical value $\langle \rm log(E_{\rm peak}/keV) \rangle = 2.70$ and standard deviation $\sigma_{E_{peak}}$ = 0.47. 
The distribution of the two photon indices $\alpha$ and $\beta$ are shown in Fig.~\ref{fig:index_short_timeint}. It is interesting to note that $\alpha$, which describes the index of the power law below \ep, has a typical value $ \langle\alpha\rangle = -0.78$ ($\sigma_\alpha$ = 0.23), i.e. consistent within 1$\sigma$ with the synchrotron value $\alpha_1^{\rm syn}-2/3$. The photon index $\beta$ of the spectral power law above \ep\ has a mean value 
$ \langle\beta\rangle = -2.59$, with $\sigma_\beta$ = 0.33.

%-------------------------------
%-------------------------------

%======================== DISCUSSION ==============================
\section{Discussion}\label{sect:discussion}

Our results show that in the majority of the brightest long GRBs detected 
by the {\it Fermi}/GBM the spectrum below the peak energy \ep\ cannot be fitted  just by a single power law;  it requires an additional break at low energies. 
This feature, unnoticed for a long time,  has been recently discovered in GRBs detected by \sw. 
\citet{gor2017a,gor2018}, in a joint \sw/XRT and BAT analysis of GRBs with prompt emission detected simultaneously by both instruments, modelled the spectrum adding a spectral break between 3\,keV and 22\,keV, and a third power law below the break energy. 
They obtained values of the photon indices below and above the break consistent with synchrotron predictions. 
This allowed them to speculate on a synchrotron origin and associate the break energy with the synchrotron cooling frequency, $\nu_{\rm c}$. 
\cite{Ravasio2018} found the same feature in the spectrum of GRB 160625B, one of the brightest GRBs ever detected by the {\it Fermi}/GBM. 
Considering the mean value of the break energy of GRB 160625B 
($E_{\rm break}\sim100$\,keV) observed in the time-resolved spectra
of 1\,s interval and
interpreting the spectrum as synchrotron emission in fast cooling regime, 
\cite{Ravasio2018} derived a limit on the value of the comoving magnetic field $B^\prime$ of the order of 
\begin{equation}
B^\prime \sim 13  \,\,\,\,  \Gamma_{2}^{-1/3} \nu_{\rm c,100 \,\rm keV}^{-1/3} \,\,\,  t_{\rm 1s}^{-2/3}\,\,\,\, \rm G~,
\end{equation}
where we  assume a typical bulk Lorentz factor of 100 and $t$ is the typical integration time of the analysed spectra.

Considering the results presented in this work, \eb\ found in the brightest \fe\ bursts is distributed in the range $\sim$\,20 -- 600\,keV (Fig.~\ref{fig:energies_long_timeres}). Since the distribution of bulk Lorentz factors $\Gamma$ (as obtained in \citealt{Ghirlanda2018}) spans two orders of magnitude from $\sim$\,20 -- 2000, we derive an estimate of the corresponding distribution of the comoving magnetic field $B^\prime\in[1,40]$\,G.

These values of the comoving magnetic field are very small compared to expectations for the typical GRB emitting region (but see e.g. \citealt{Kumar2007, Zhang2009, Zhang2011b} for a Poyting flux dominated outflow where a low magnetic field can be achieved at large radii).
Therefore, while our results positively solve the issue of the inconsistency of observed spectra and synchrotron radiation, they open a new challenging question: within the standard GRB model and synchrotron theory, having the cooling break at a few hundred  keV implies that the magnetic field of the emission region is very small, which is at odds with the MGauss value expected according to the standard model
(for quasi-constant jet Poynting flux).
The problem then shifts on the search for a mechanism that can justify such a low magnetic field in the emission region.

%%%%%%%%%%%%%%%%%%%%%% short
The case of short GRBs is even more problematic.
As shown in Sec.~\ref{subsec:short}, 
short GRBs have a single power law below the $\nu F_{\nu}$ peak, characterised  by a hard photon index:  $\alpha = -0.78 \pm 0.23$. 
The power law between \eb\ and \ep, namely the one with photon index $\alpha_2 = -1.5$, seems to be missing in short GRBs.
This suggests that electrons do not cool efficiently in short GRBs, 
implying that the magnetic field is even smaller than in long GRBs,
raising an
efficiency problem which is difficult to explain within the scenario of the standard model.

A self-consistent 
picture for the prompt emission mechanism should explain, among other things, 
i) spectra with two breaks,  which we have found in this work for bright long GRBs; 
ii) the variability of the prompt emission; and 
iii) the huge amount of energy radiated during the prompt. 
We cautiously note, however, that these considerations are drawn under 
the hypothesis of the synchrotron process: if the emission is not due to 
this process, then  some other radiation mechanism will need to be invoked  to explain the current findings.

%===================================== CONCLUSIONS =====================================
\section{Conclusions}\label{sect:conclusions}

In this work we presented the spectral analysis of the brightest ten long and ten short GRBs detected by {\it Fermi}/GBM in 10 years of activity. 
We systematically fitted two empirical functions to the spectra: a smoothly broken power law (SBPL) and a double smoothly broken power law (2SBPL). The reason for testing a 2SBPL model was to identify the possible presence of two characteristic energies in the prompt emission spectra: the usual peak energy \ep\ and a spectral break \eb\ at lower energies, recently identified in a sample of \sw\ bursts \citep{gor2017a,gor2018} and in one \fe/GBM bright burst \citep{Ravasio2018}.

For long GRBs, the time-integrated analysis shows that in eight of the ten brightest GBM GRBs, the standard empirical fitting function (SBPL) fails to provide an acceptable fit: the data require an additional spectral break \eb, located between $\sim$\,10\,keV and 300\,keV. 
For these eight GRBs we also performed a time-resolved analysis, finding that $\sim70\%$ of the time-resolved spectra also show strong evidence of a low-energy spectral break. 
For this sample of time-resolved spectra from eight bright long GRBs, the log-normal distributions of \eb\ and \ep\ are centred around the mean values $\langle \log(E_{\rm break}/{\rm keV})\rangle=2.00 \pm 0.34$ and $\langle \log(E_{\rm peak}/{\rm keV}) \rangle=3.00 \pm 0.26$. 
The spectrum below \eb\ is nicely described by a power law. The photon indices of the power laws below \eb\ and between \eb\ and \ep\ are, respectively, $\langle\alpha_1\rangle = -0.58 \pm 0.16$ and 
$\langle\alpha_2\rangle = -1.52 \pm 0.20$, remarkably consistent with the predicted values for synchrotron emission in a marginally fast cooling regime (for an example of a typical spectrum, see Fig.~\ref{fig:spettro180720}).

The remaining time-resolved spectra ($\sim$\,30\%) are best fitted by a simple SBPL, i.e. the improvement in   $\chi^2$ found by fitting a 2SBPL to the data has a significance in that is smaller than the threshold value of $3\sigma$.
In these cases, one power law is sufficient to model the spectra below \ep, and its typical value is $\langle\alpha\rangle=-1.02\pm0.19$. 
Interestingly, this value lies between the values of 
$\alpha_1$ and $\alpha_2$, as shown in 
Fig.~\ref{fig:distrib_index_long_timeres}. 
Speculating that most of the spectra present a break below \ep, the value of $\alpha$ can be understood as a sort of average value between  $\alpha_1$ and $\alpha_2$: these are indeed asymptotic values that can be reached if \eb\ and \ep\ are far from each other.
Moreover, we note that when the model is a simple SBPL, the fit tends to place \ep\ at a smaller energy, thus also resulting in a softer $\beta$ (see Fig.~\ref{fig:distrib_index_long_timeres}).

For short GRBs, none of the time-integrated spectra of the ten brightest events shows a break at low energies. 
The best fit model is always a SBPL, and the distribution of the index below 
\ep\ is centred at $ \langle\alpha\rangle = -0.78 \pm 0.23$.
As for $\alpha_1$ in long GRBs, this value is consistent within $1\sigma$ with the low-energy (below $h\nu_{\rm c}$) synchrotron photon index 
$\alpha_1^{\rm syn}$.
In a synchrotron interpretation, this implies that in short GRBs $\nu_{\rm c}$ is even closer to $\nu_{\rm m}$ and the power law between \eb\ and \ep, namely the one with asymptotic spectral index $\alpha_2^{\rm syn}=-1.5$, is missing.
 
In both  long and  short GRBs we find that the hard spectrum below \eb\ and its photon index suggest a synchrotron origin for the observed GRB prompt spectra.
As discussed in \S \ref{sect:discussion}, assuming that the observed \eb\ corresponds to the synchrotron cooling frequency, the implied magnetic field strength in the emitting region is small (between 1 and 40\,G in the comoving frame), i.e. orders of magnitudes smaller than expected for a dissipation region located at $\sim10^{13-14}$\,cm from the central engine.

If electrons really cool over a relatively long time, to give rise to the observed
low-energy slope, they are emitting in a relatively small magnetic field,
at odds with the expectations that the magnetic field plays a major role 
to power and launch the GRB jet.
This requires a major revision of the standard GRB model.

%==========================================================================
\begin{acknowledgements}
M.E.R. is thankful to the Observatory of Brera for the kind hospitality. INAF-Prin 2017  funding is acknowledged. 
L.N. acknowledges funding from the European Union's Horizon 2020 Research and Innovation programme under the Marie Sk\l odowska-Curie grant agreement n.\,664931.
This research has made use of data obtained through the High Energy Astrophysics Science Archive Research Center Online Service provided by the NASA/Goddard Space Flight Center, and specifically this work has made use of public \fe-GBM data. We also would like to thank for support the implementing agreement ASI-INAF n.2017-14-H.0.
\end{acknowledgements}

\bibliographystyle{aa} 
\bibliography{references} 

\begin{appendix}
\section{Tables}\label{app:tables}

\begin{table*}
\small
\centering 
\caption{Best fit parameters inferred from the time-integrated analysis of the ten long GRBs analysed in this work. 
In bold  are flagged those GRBs with a statistically significant spectral break \eb\ in the low-energy part of their spectrum.
The table lists the GRB name, the time interval over which the spectrum has been accumulated, and the results from the spectral analysis: best fit normalisation (see \citealt{Ravasio2018}), photon index $\alpha_1$ (or $\alpha$ when the best fit model is a SBPL), break energy \eb\ (only if the best fit model is a 2SBPL), photon index $\alpha_2$ (only if the best fit model is a 2SBPL), peak energy \ep, photon index $\beta$, total $\chi^2$ and degrees of freedom (dof), and the significance of the improvement of the fit from a SBPL to a 2SBPL (estimated according to the $F$-test).
}
\label{tab:params_timeint_long}
\begin{tabular}{cccccccccc}
\hline\hline
  \multicolumn{1}{c}{Name} &
  \multicolumn{1}{c}{Time interval} &
  \multicolumn{1}{c}{Norm} &
  \multicolumn{1}{c}{$\alpha_1$ ($\alpha$)} &
  \multicolumn{1}{c}{$E_{\rm break}$} &
  \multicolumn{1}{c}{$\alpha_2$} &
  \multicolumn{1}{c}{$E_{\rm peak}$} &
  \multicolumn{1}{c}{$\beta$} &
  \multicolumn{1}{c}{$\chi^2$/dof} &
  \multicolumn{1}{c}{$F_{\rm test}$} \\
  & [s] & [$\rm ph / s\, cm^{2}\, keV$] & & [keV] & & [keV] & & & \\
\hline
{\bf 171010} & [0.003 - 100.35 s] & $ { 0.12 }_{- 0.04 }^{+ 0.04 }$ & ${ +1.16 }_{- 0.13 }^{+ 0.13 }$ & ${ 12.39 }_{- 0.13 }^{+ 0.13 }$ & ${ -1.4 }_{- 0.01 }^{+ 0.01 }$ & ${ 182.2 }_{- 1.8 }^{+ 1.8 }$ & ${ -2.7 }_{- 0.02 }^{+ 0.02 }$ & 829.22 / 335 & $>8\sigma$ \\
{\bf 160625} & [187.43 - 212.00 s] & $ { 4.58 }_{- 0.15 }^{+ 0.15 }$ & ${- 0.56 }_{- 0.01 }^{+ 0.01 }$ & ${ 119.9 }_{- 3.79 }^{+ 3.79 }$ & ${ -1.7 }_{- 0.03 }^{+ 0.03 }$ & ${ 646.5 }_{- 18.0 }^{+ 18.0 }$ & ${ -2.67 }_{- 0.03 }^{+ 0.03 }$ & 638.55 / 342 & $>8\sigma$ \\
{\bf 160821} & [117.76 - 154.63 s] & $ { 9.08 }_{- 0.51 }^{+ 0.48 }$ & ${ -0.87 }_{- 0.02 }^{+ 0.02 }$ & ${ 158.4 }_{- 22.3 }^{+ 21.4 }$ & ${ -1.59 }_{- 0.05 }^{+ 0.05 }$ & ${ 1295.0 }_{- 50.0 }^{+ 55.8 }$ & ${ -2.61 }_{- 0.05 }^{+ 0.05 }$ & 411.91 / 226 & $>8\sigma$ \\
{\bf 170409} & [17.66 - 116.99 s] & $ { 1.68 }_{- 0.07 }^{+ 0.07 }$ & ${ -0.88 }_{- 0.01 }^{+ 0.01 }$ & ${ 315.3 }_{- 24.3 }^{+ 24.1 }$ & ${ -1.78 }_{- 0.05 }^{+ 0.05 }$ & ${ 1156.0 }_{- 81.5 }^{+ 93.4 }$ & ${ -3.39 }_{- 0.15 }^{+ 0.18 }$ & 527.62 / 347 & $>8\sigma$ \\
{\bf 180720} & [-1.02 - 56.32 s] & $ { 4.19 }_{- 0.77 }^{+ 0.85 }$ & ${ -0.73 }_{- 0.09 }^{+ 0.08 }$ & ${ 38.12 }_{- 8.32 }^{+ 12.6 }$ & ${ -1.48 }_{- 0.05 }^{+ 0.06 }$ & ${ 774.8 }_{- 36.9 }^{+ 50.0 }$ & ${ -2.61 }_{- 0.05 }^{+ 0.07 }$ & 589.06 / 343 & $>8\sigma$ \\
{\bf 171227} & [0.003 - 58.24 s] & $ { 1.92 }_{- 0.14 }^{+ 0.14 }$ & ${ -0.75 }_{- 0.02 }^{+ 0.02 }$ & ${ 153.3 }_{- 14.5 }^{+ 14.4 }$ & ${ -1.68 }_{- 0.05 }^{+ 0.05 }$ & ${ 1064.0 }_{- 69.8 }^{+ 81.5 }$ & ${ -2.98 }_{- 0.1 }^{+ 0.11 }$ & 466.24 / 344 & $>8\sigma$ \\
{\bf 090618} & [0.003 - 161.28 s] & $ { 2.27 }_{- 1.1 }^{+ 0.43 }$ & ${ -0.19 }_{- 2.83 }^{+ 0.07 }$ & ${ 7.75 }_{- 0.81 }^{+ 1.26 }$ & ${ -1.5 }_{- 0.03 }^{+ 0.02 }$ & ${ 157.2 }_{- 6.27 }^{+ 6.8 }$ & ${ -2.87 }_{- 0.1 }^{+ 0.1 }$ & 339.3 / 231 & $ 4.1 \sigma$ \\
{100724} & [-5.12 - 137.22 s] & $ { 1.26 }_{- 0.05 }^{+ 0.05 }$ & ${ -0.87 }_{- 0.01 }^{+ 0.01 }$ & - & - & ${ 659.9 }_{- 75.0 }^{+ 132.0 }$ & ${ -2.05 }_{- 0.02 }^{+ 0.02 }$ & 444.62 / 339 & $ 1.0 \sigma$ \\
{130606} & [-3.07 - 70.66 s] & $ { 10.17 }_{- 0.36 }^{+ 0.37 }$ & ${ -1.19 }_{- 0.01 }^{+ 0.01 }$ & - & - & ${ 600.7 }_{- 52.4 }^{+ 69.9 }$ & ${ -2.11 }_{- 0.02 }^{+ 0.02 }$ & 483.38 / 333 & $ 2.0 \sigma$ \\
{\bf 101014} & [0.003 - 466.44 s] & $ { 0.6 }_{- 0.1 }^{+ 0.04 }$ & ${ -0.05 }_{- 0.0 }^{+ 0.03 }$ & ${ 10.99 }_{- 0.91 }^{+ 0.88 }$ & ${ -1.38 }_{- 0.04 }^{+ 0.03 }$ & ${ 221.7 }_{- 14.9 }^{+ 18.3 }$ & ${ -2.35 }_{- 0.07 }^{+ 0.08 }$ & 488.14 / 342 & $ 7.4 \sigma$ \\
\hline\end{tabular}
\end{table*}

\begin{table*}
\small
\centering 
\caption{Average results of the time-resolved analysis for the seven long GRBs that display a low-energy break \eb\ in their time-integrated spectrum (see GRBs  in bold  in Table~\ref{tab:longGRB}; GRB~171010 is not included, because all time-resolved spectra have \eb$<20$\,keV).
For each GRB, the table lists the mean value and standard deviation $\sigma$ of the best fit parameter distribution inferred from the time-resolved analysis for spectra with best fit model given by a 2SBPL and with $E_{\rm break}>20$\,keV. 
The second column gives the number of spectra satisfying these conditions
over the total number of spectra analysed. 
} 
\label{tab:params_timeres_long}
\begin{tabular}{cccccccc}
\hline\hline
  \multicolumn{1}{c}{Name} &
  \multicolumn{1}{c}{$\#$ of spectra} &
  \multicolumn{1}{c}{$\left\langle \alpha_1 \right\rangle$} &
  \multicolumn{1}{c}{$\left\langle \alpha_2 \right\rangle$} &
  \multicolumn{1}{c}{$\left\langle E_{\rm break} \right\rangle$} &
  \multicolumn{1}{c}{$\left\langle E_{\rm peak} \right\rangle$} &
  \multicolumn{1}{c}{$\left\langle \beta \right\rangle$} \\
  & & & & [keV] & [keV] & \\
\hline
  160625 & 18/24 & $-0.51$ (0.08) & $-1.62$ (0.15) & $110.66$ (22.65) & $805.98$ (668.66) & $-2.79$ (0.24) \\
  160821 & 17/27 & $-0.74$ (0.15) & $-1.51$ (0.17) & $133.49$ (94.49) & $1643.16$ (745.55) & $-2.62$ (0.17) \\
  170409 & 10/14 & $-0.62$ (0.0) & $-1.66$ (0.25) & $334.60$ (141.47) & $1304.06$ (656.459) & $-3.47$ (0.37) \\
  180720 & 15/24 & $-0.54$ (0.14) & $-1.41$ (0.15) & $55.67$ (39.43) & $1093.26$ (481.31) & $-2.58$ (0.20) \\
  171227 & 10/11 & $-0.47$ (0.10) & $-1.43$ (0.13) & $134.29$ (36.39) & $1212.26$ (348.01) & $-2.90$ (0.24) \\
  $090618$ & 1/18 & $-0.83$ (0.10) & $-1.77$ (0.20) & $118.56$ (50.36) & $550.05$ (93.29) & $-3.57$ (0.55) \\
  $101014$ & 3/17 & $-0.32$ (0.23) & $-1.32$ (0.11) & $36.31$ (14.36) & $862.77$ (113.23) & $-2.62$ (0.16) \\
  \hline
  Mean values: & & $-0.58$ (0.16) & $-1.52$ (0.20) & $135.31$ (112.15) & $1177.59$ (679.28) & $-2.81$ (0.37)\\
\end{tabular}
\end{table*}

%---------------------------------------------------------------
\begin{table*}
\centering 
\caption{Best fit parameters for the time-integrated analysis of the ten  brightest short GRBs analysed in this work. The best fit model is always a SBPL.}
\label{tab:params_timeint_short}
\small
\begin{tabular}{cccccccc}
\hline\hline
  \multicolumn{1}{c}{Name} &
  \multicolumn{1}{c}{Time interval} &
  \multicolumn{1}{c}{Norm} &
  \multicolumn{1}{c}{$\alpha$} &
  \multicolumn{1}{c}{$E_{\rm peak}$} &
  \multicolumn{1}{c}{$\beta$} &
  \multicolumn{1}{c}{$\chi^2$/dof} &
  \multicolumn{1}{c}{Prob}  \\
& [s] & [$\rm ph / s\, cm^{2}\, keV$] & & [keV] & & \\
\hline
  170206 & [-0.128 - 1.664 s] & $1.41_{-0.20}^{+0.21}$ & $-0.62_{-0.04}^{+0.03}$ & $280.4_{-13.3}^{+18.1}$ & $-2.38_{-0.08}^{+0.06}$ & 361.63 / 346 &  0.2707 \\
  120323 & [-0.064 - 0.576 s] & $98.11_{-14.1}^{+22.6}$ & $-1.04_{-0.06}^{+0.08}$ & $109.5_{-4.87}^{+20.5}$ & $-2.11_{-0.05}^{+0.03}$ & 372.44 / 353 & 0.2286 \\
  090227 & [-0.064 - 0.256 s] & $1.68_{-0.23}^{+0.19}$ & $-0.60_{-0.03}^{+0.02}$ & $1576.0_{-67.1}^{+67.3}$ & $-2.82_{-0.20}^{+0.06}$ & 358.17 / 349 & 0.3559 \\
  150819 & [-0.064 - 1.152 s] & $25.21_{-2.52}^{+2.3}$ & $-1.24_{-0.03}^{+0.02}$ & $595.3_{-73.8}^{+188.0}$ & $-2.41_{-0.19}^{+0.16}$ & 349.53 / 347 & 0.4518 \\
  170127 & [-0.064 - 0.256 s] & $0.78_{-0.39}^{+0.19}$ & $-0.47_{-0.11}^{+0.05}$ & $755.5_{-56.0}^{+23.4}$ & $-3.29_{-0.59}^{+0.06}$ & 312.4 / 348  & 0.9151 \\
  120624 & [0.000 - 0.320 s] & $5.1_{-0.62}^{+0.62}$ & $-0.83_{-0.02}^{+0.02}$ & $2892.0_{-205.0}^{+299.0}$ & $-2.49_{-0.17}^{+0.12}$ & 344.93 / 345 & 0.491 \\
  130701 & [-0.064 - 1.600 s] & $0.46_{-0.11}^{+0.12}$ & $-0.69_{-0.06}^{+0.04}$ & $892.3_{-74.0}^{+90.9}$ & $-2.68_{-0.26}^{+0.21}$ & 324.75 / 347 & 0.7989 \\
  130504 & [-0.032 - 0.384 s] & $0.87_{-0.17}^{+0.15}$ & $-0.57_{-0.04}^{+0.03}$ & $1033.0_{-46.9}^{+74.9}$ & $-2.79_{-0.25}^{+0.10}$ & 375.6 / 352 & 0.1853 \\
  090228 & [-0.064 - 0.512 s] & $2.29_{-0.33}^{+0.32}$ & $-0.76_{-0.03}^{+0.03}$ & $663.7_{-38.0}^{+51.3}$ & $-2.87_{-0.28}^{+0.17}$ & 346.76 / 349 & 0.5238 \\
  
\hline\end{tabular}
\end{table*}

%---------------------------------------------------------------
\section{Spectra with $E_{\rm break}<20$\,keV}\label{appendix:Ebreaklow}

\begin{figure}[h]
    \centering
    \includegraphics[scale=0.43]{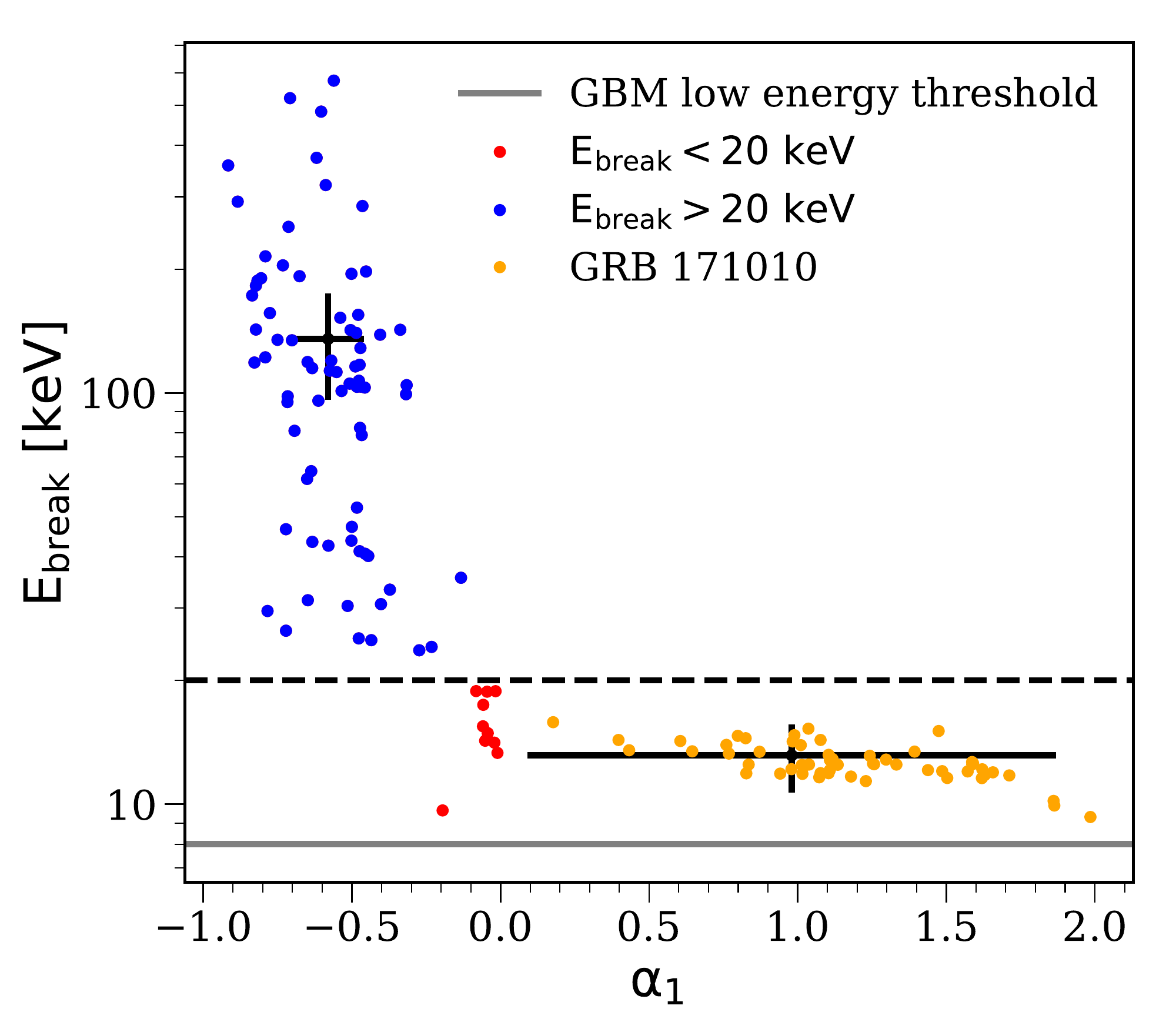}
    \caption{Relation between the break energy \eb\ and the index $\alpha_1$ of the power law describing the spectrum below \eb. The sample includes all time-resolved spectra of the ten long GRBs analysed in this work. 
    The dashed black horizontal line indicates a difference in behaviour, with spectra below this line having very hard and ill-constrained best fit values of $\alpha_1$. All the parameter distributions and their mean values and standard deviations presented in this work rely only on spectra with \eb\ larger than this threshold value (blue points). The majority ($\sim 85\%$) of the spectra with $E_{\rm break}<20$\,keV (red points) belong to one specific GRB, namely GRB\,171010 (orange points). The mean values of \eb\ and $\alpha_1$ (along with their average errors) are represented for each sample with the solid black lines. The low-energy threshold of the GBM NaI detectors is shown with a grey solid line.}
    \label{fig:Eb_vs_a1}
\end{figure}

In this section we discuss spectra where the 2SBPL parameter \eb\ has a best fit value close to the low-energy threshold of sensitivity of the GBM ($E_{\rm th}\sim 8$\,keV).
In these spectra, $\alpha_1$ behaves very differently  to typical $\alpha_1$ values inferred for all the other spectra.
Figure~\ref{fig:Eb_vs_a1} shows \eb\ versus $\alpha_1$ for the full sample of long GRBs (time-resolved analysis). 
A sudden change in the location of points is visible at low \eb, with a  well-defined separation at \eb\,$\sim$\,20\,keV (dashed horizontal line).
In particular, spectra with \eb\,$<20$\,keV (red and orange symbols) have considerably harder values  of the low-energy photon index (i.e. $\alpha_1>-0.2$).
The uncertainty on these values is large, as shown by the black cross plotted on top of the orange/yellow points, which represents the average errors on the two parameters represented. 
When the break energy is $>\sim$20\,keV, the distribution of $\alpha_1$ is completely different, with almost no overlapping between the two distributions. Also, the distribution is narrower and the typical error is smaller (black cross plotted on top of the blue points). 

The peculiar distribution of the points in the \eb-$\alpha_1$ plane strongly suggests an instrumental effect at the origin of the hard values derived when \eb$<20$\,keV. 
The low-energy edge of sensitivity of the GBM is $E_{\rm th}\sim 8$\,keV (solid grey line in Fig.~\ref{fig:Eb_vs_a1}), implying that when \eb\ is below 20\,keV, a few channels are available for the determination of $\alpha_1$. 
Even though this is most certainly true, it is less evident why  $\alpha_1$ should be systematically overestimated in these
cases.

We also note that $\sim$\,85\% of the time resolved spectra with \eb\,$<20$\,keV  belong to a single GRB, i.e. 171010 (orange symbols in Fig.~\ref{fig:Eb_vs_a1}, see also \citealt{Chand2018}).
\sw/XRT data (as in the GRBs analysed  by \citealt{gor2017a}) would be of paramount importance in cases like this one to better characterise the low-energy photon index of GRB\,171010 and other similar cases. Unfortunately, for GRB\,171010 there are no {\it Swift}/XRT data simultaneous to the GBM data. 

%%%%%%%%%%%%%%%%%%%%%%%%%%%%%%%%%%%%

%---------------------------------------------------------------

\section{Spectral evolution of individual GRBs}\label{appendix:timeevolutionpanels}
In this section, we show the light curve and temporal evolution of the best fit spectral parameters for each GRB with break energy identified in the time-integrated spectrum (i.e. eight long GRBs).

\begin{figure*}[h]
    \centering
    \includegraphics[scale=0.25]{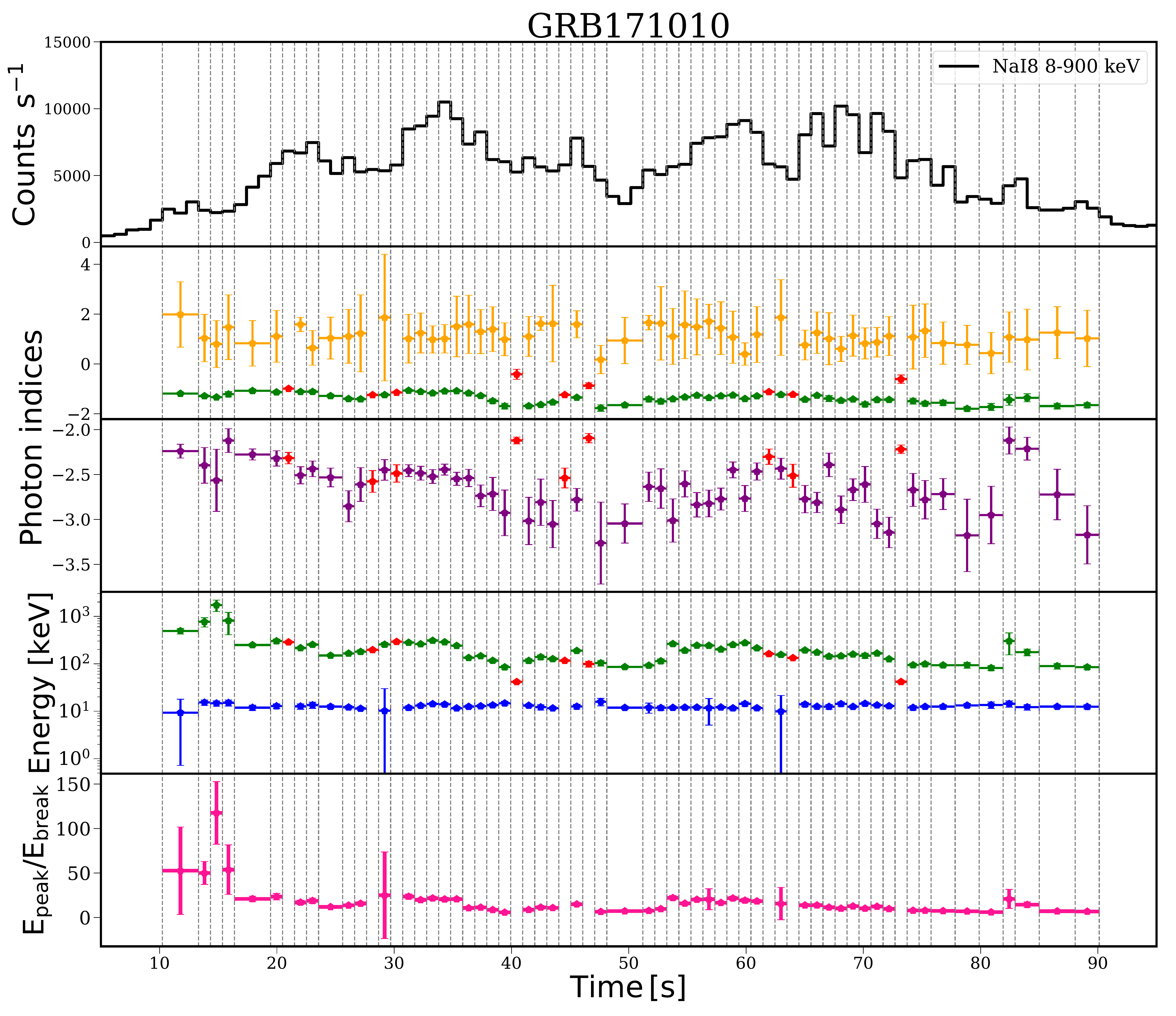}
    \includegraphics[scale=0.25]{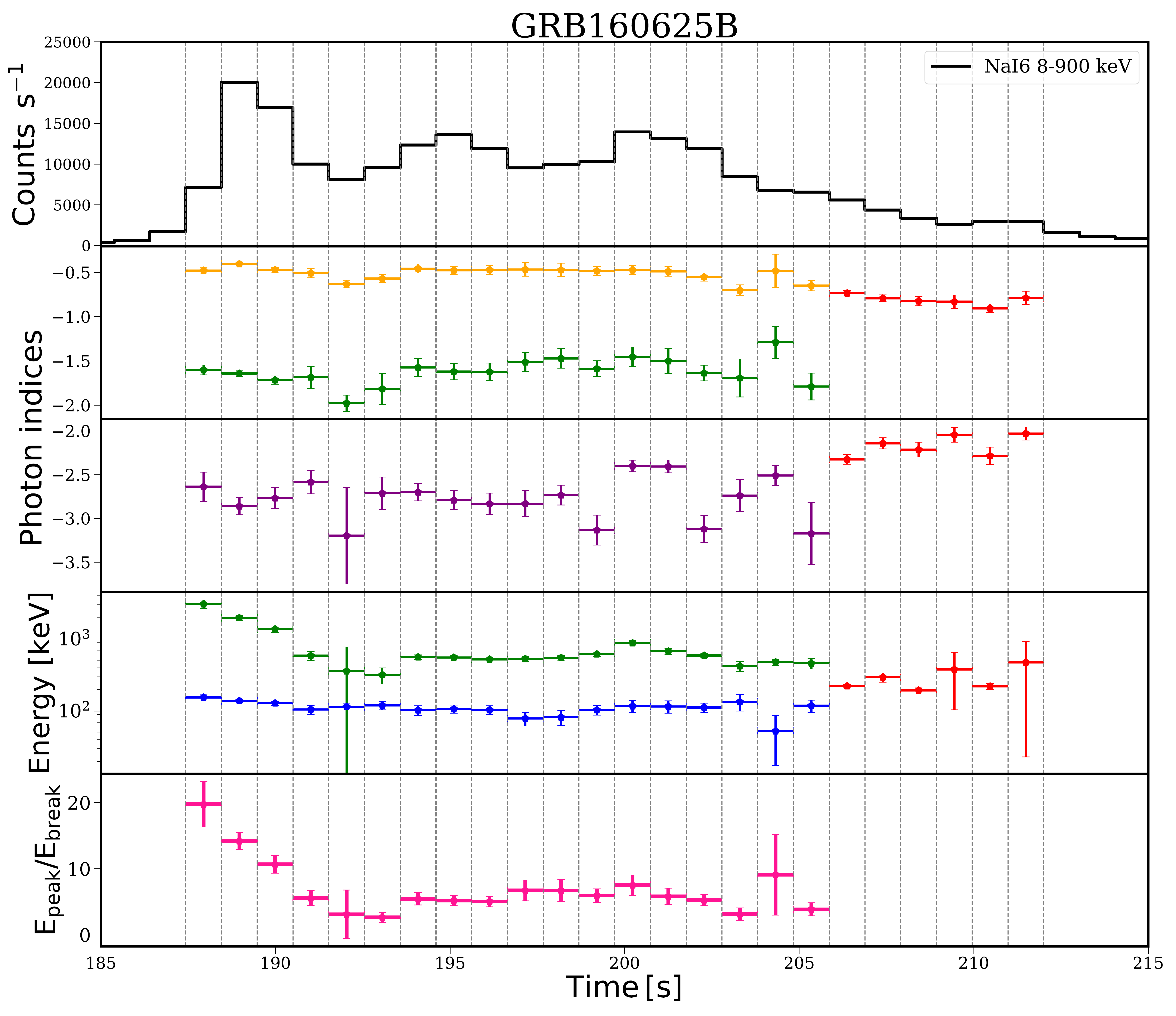}
    \label{fig:timeevolutionpanels}
    \end{figure*}
    
\begin{figure*}[h]
    \centering
    \includegraphics[scale=0.25]{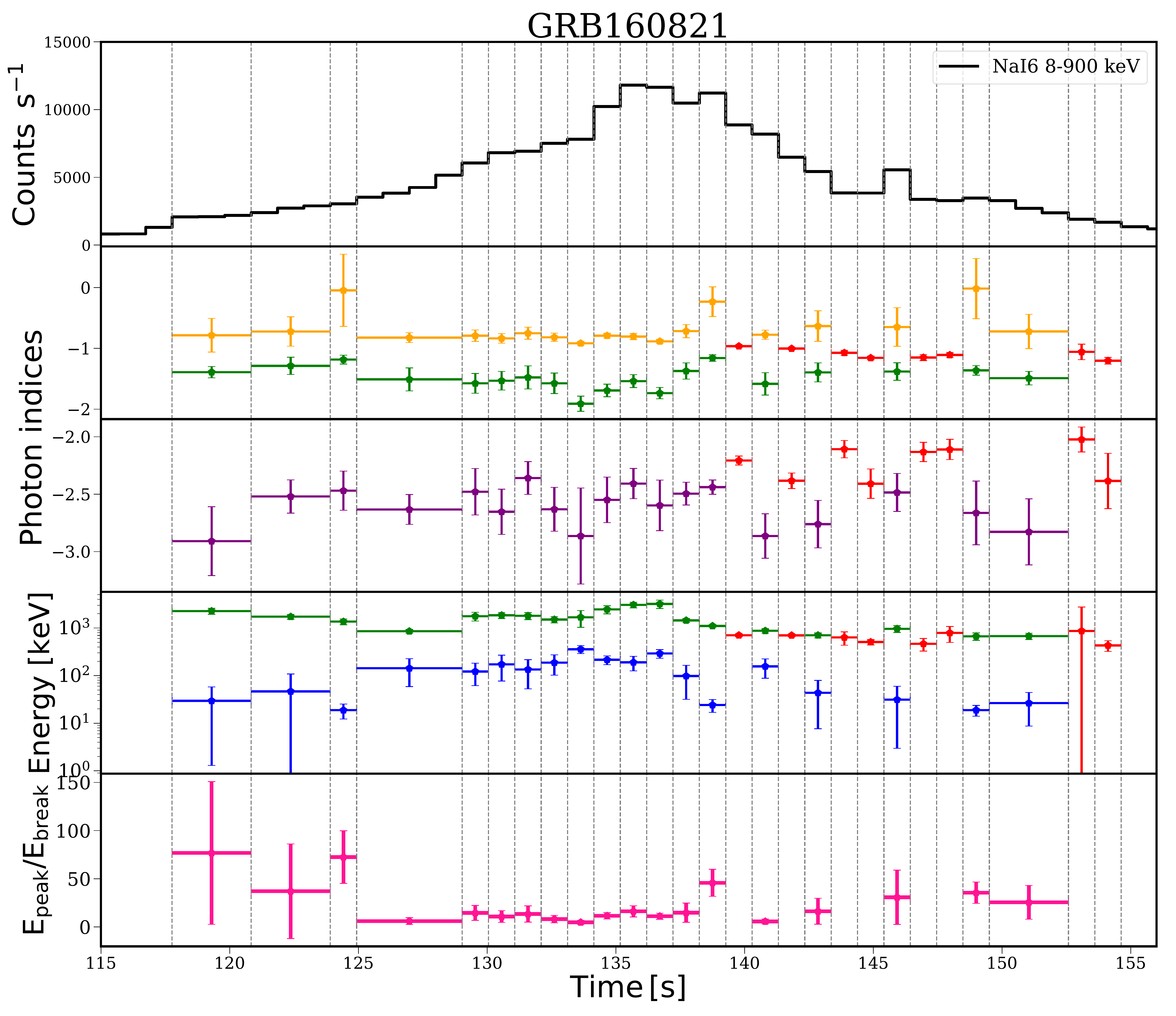}
\end{figure*}

\begin{figure*}[h]
    \centering
    \includegraphics[scale=0.25]{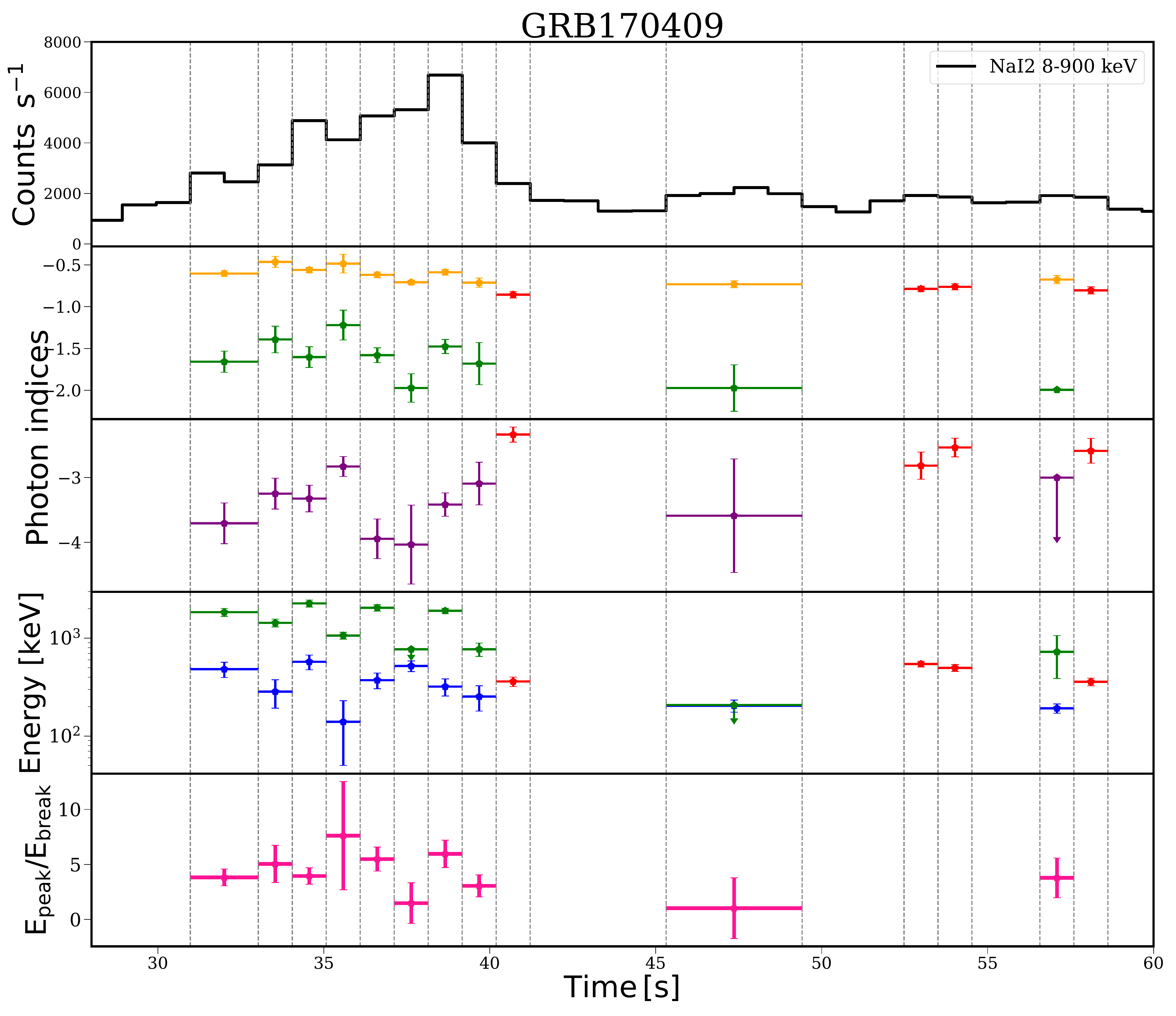}
\end{figure*}

\begin{figure*}[h]
    \centering
    \includegraphics[scale=0.25]{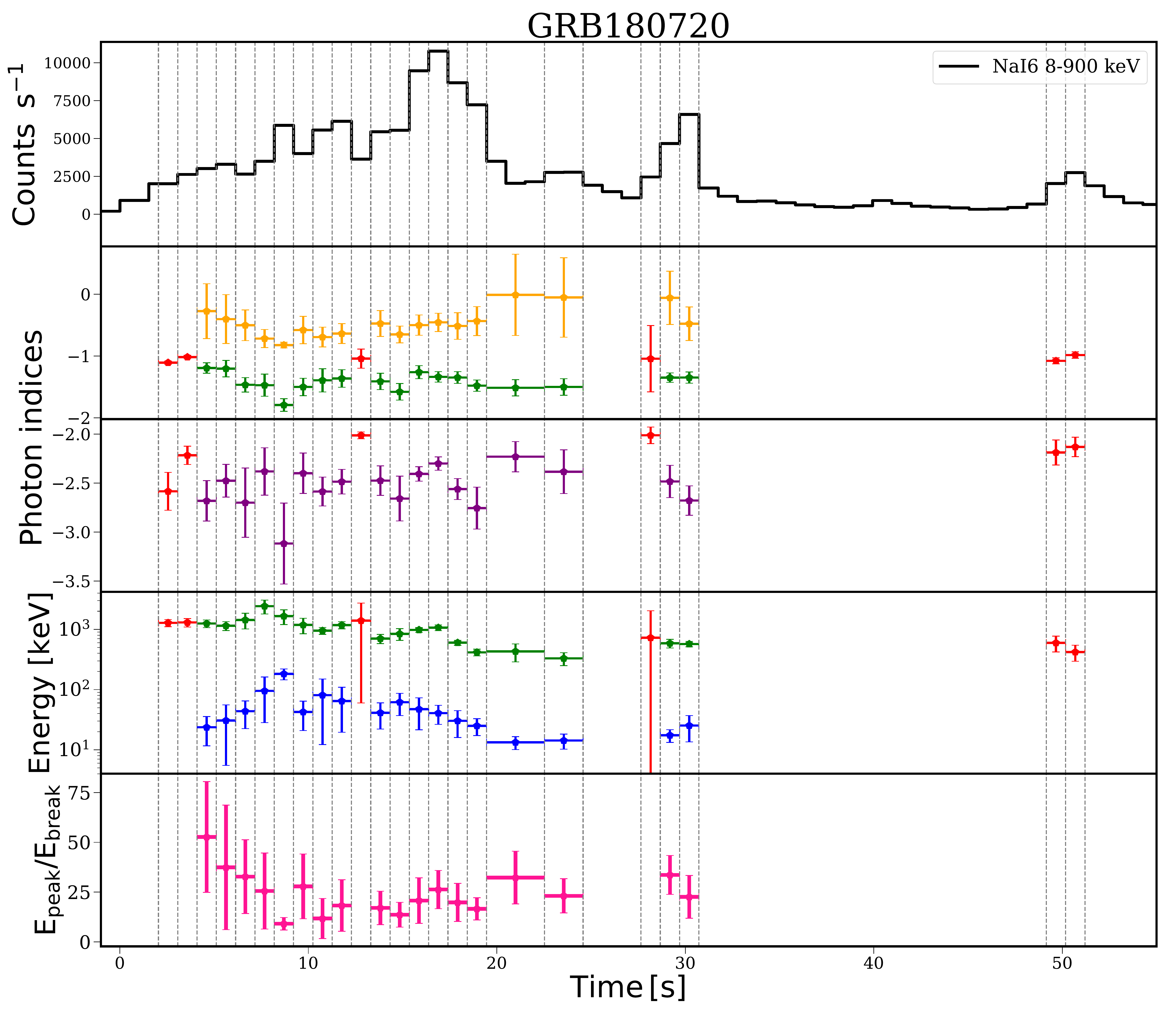}
\end{figure*}

\begin{figure*}[h]
    \centering
    \includegraphics[scale=0.25]{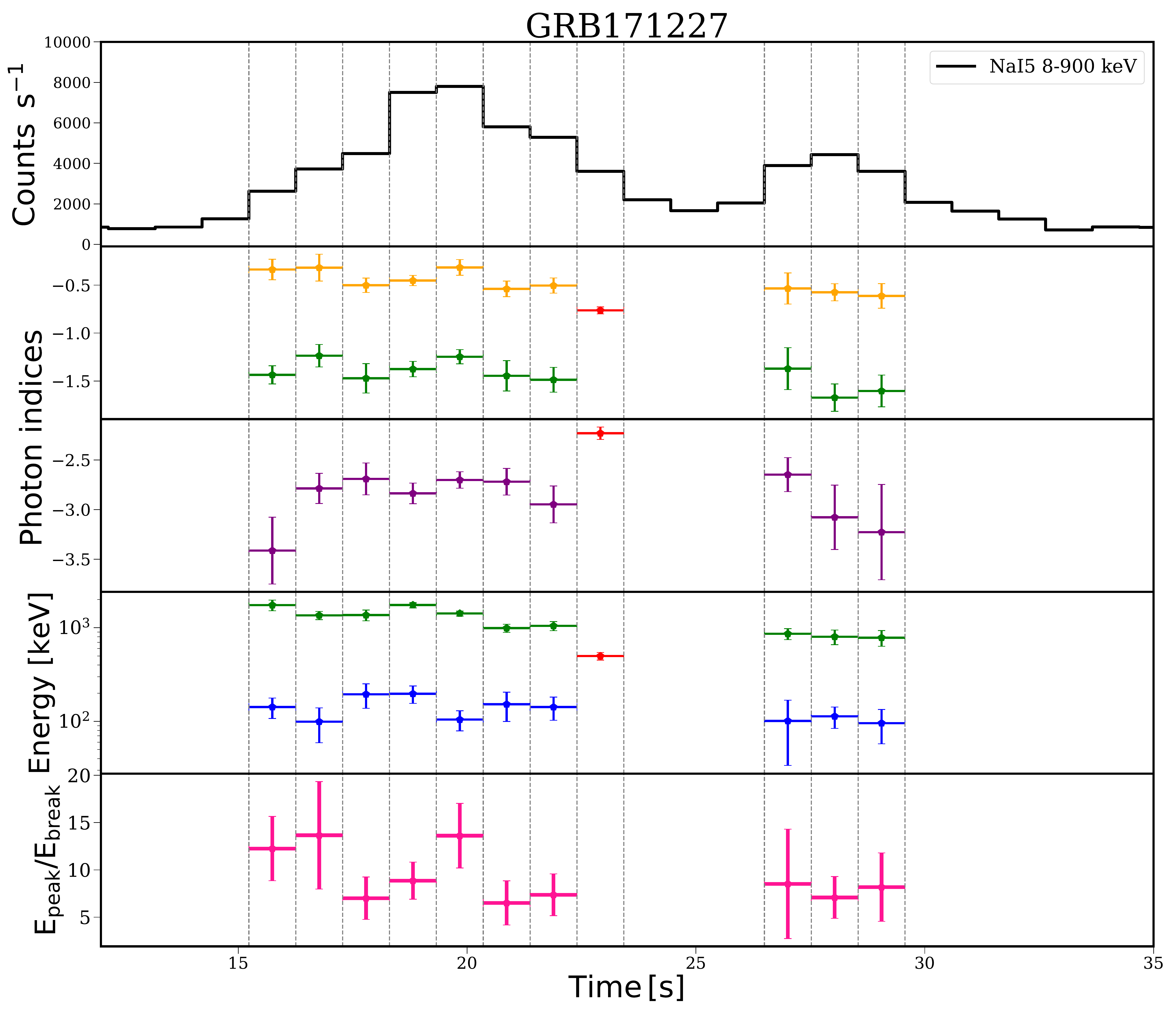}
\end{figure*}

\begin{figure*}[h]
    \centering
    \includegraphics[scale=0.25]{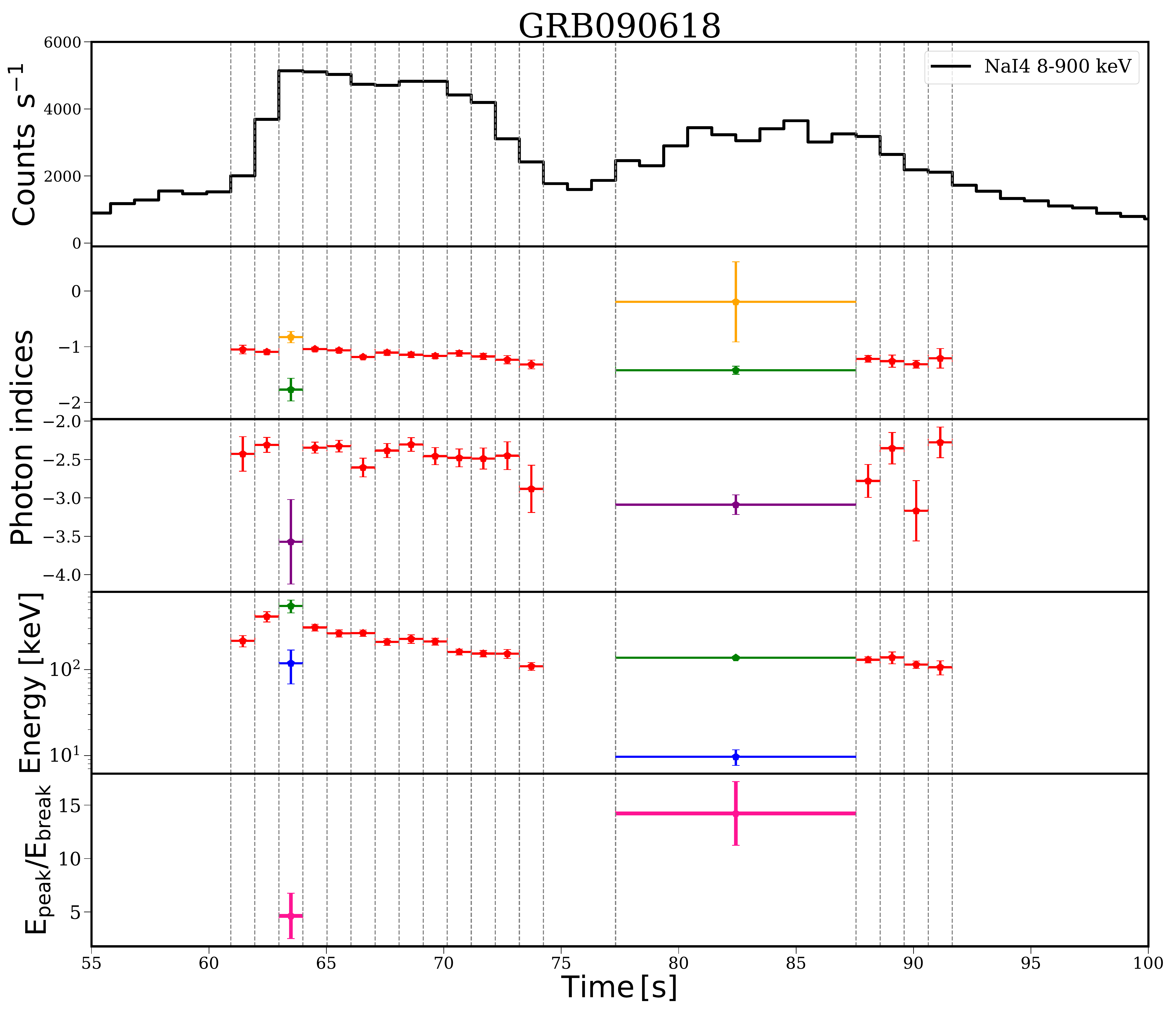}
\end{figure*}

\begin{figure*}[h]
    \centering
    \includegraphics[scale=0.25]{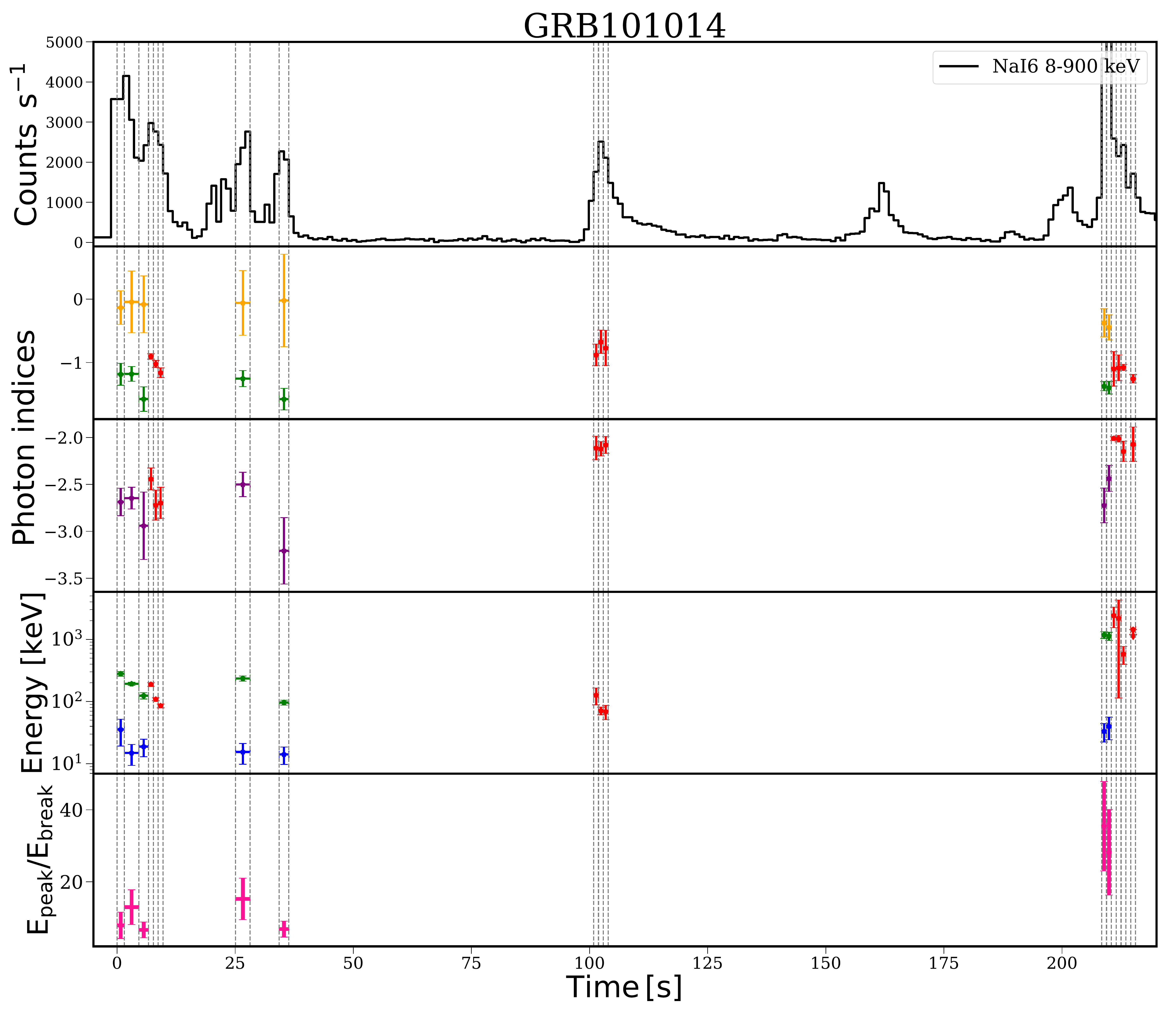}
     \caption{These figures show the evolution of the best fit spectral parameters for each GRB displaying a low-energy break. The first panel shows the light curve of the GRB, in the energy range 8-900 keV. The vertical lines mark the time bins selected for the time-resolved analysis. The panels below show the temporal evolution of all best fit parameters (of the 2SBPL or SBPL function, according to which model fits best the spectrum of the time bin). While all the parameters of the SBPL fit are shown as red points, different colours have been used to represent the parameters of the 2SBPL fit. In particular, from top to bottom, in the second panel there are the photon indices $\alpha_{1}$ (yellow points) and $\alpha_{2}$ (green points) of the 2SBPL function, and $\alpha$ (red points) of the SBPL function. In the third panel there are the two photon indices $\beta$ (in purple for the 2SBPL function and in red for  SBPL). The fourth panel shows $E_{\rm break}$ (blue points) and $E_{\rm peak}$ (green points) for  2SBPL, and $E_{\rm peak}$ for  SBPL (red points). The bottom panel shows the ratio $E_{\rm peak} / E_{\rm break}$.}
\end{figure*}

\end{appendix}

%-------------------------------------------------------------------

\end{document}